\documentclass[12pt]{article}

\usepackage{geometry, graphics, graphicx, epstopdf, xcolor, color, amssymb, amsmath, enumerate, multirow, setspace,natbib}
\usepackage{mathtools, lineno, setspace, xr, authblk}
\usepackage{soul}
\usepackage{hyperref}
\hypersetup{colorlinks,urlcolor=blue,linkcolor=red,citecolor=black}

\usepackage{caption}
\DeclareCaptionLabelFormat{nospace}{#1#2}
\usepackage{newfloat}

\DeclareFloatingEnvironment[
    name=Extended Data Fig.,
    listname={List of Extended Data Figures}
]{extdatafigure}

\date{}
\title{Climatic Phase Transitions Unravel the Onset and Withdrawal of Indian Monsoon}
\author[1, 2]{Yogenraj Patil}
\author[1, 2]{Gaurav Chopra}
\author[1, 2]{Shruti Tandon}
\author[3]{B. N. Goswami}
\author[1, 2,*]{R. I. Sujith}
\affil[1]{Department of Aerospace Engineering, Indian Institute of Technology Madras, Chennai-600036, India}
\affil[2]{Centre of Excellence for studying Critical Transitions in Complex Systems, Indian Institute of Technology Madras, Chennai-60036, India}
\affil[3]{Department of Physics, Gauhati University, Guwahati-781014, India}
\affil[*]{\textit{Corresponding author, email: sujith@iitm.ac.in}}
\begin{document}
\maketitle




\newpage
\begin{abstract}
The livelihood and food security of more than a billion people depend on the Indian monsoon (IM)\citep{kathayat2017indian}. Yet, a universal definition of the large-scale season and progress of IM is missing. Even though IM is a planetary-scale convectively coupled system arising largely from seasonal migration of the Intertropical Convergence Zone (ITCZ)\citep{webster1998monsoons, gill1980some, gadgil2003indian, goswami2017dynamics}, the definitions of its onset and progression are based on local weather observations, making them practically inutile due to the detection of bogus onsets. Using climate networks, we show that small-scale clusters of locally defined rainfall onsets coalesce through two abrupt climatic phase transitions defining large-scale monsoon onsets over Northeast India and the Indian peninsula, respectively. These abrupt transitions are interspersed with continuous growth of clusters. Breaking the conventional wisdom that IM starts from southern peninsula and expands northward and westward, we unveil that IM starts from Northeast India and expands westward and northward, covering the entire country. We show that the large-scale monsoon onset over the Indian peninsula is critically dependent on the characteristics of monsoon onset over Northeast India. Unlike existing definitions\citep{pai2020normal, misra2018local, xavier2007objective, walker2016onset, fasullo2003hydrological}, a rapid and consistent northward propagation of rainfall establishing the ITCZ manifests after our network-based onset dates. Thus, our definition captures the IM onset better than the existing definitions. 
\end{abstract}

\newpage
\section*{Introduction}
The Indian Monsoon (IM) is an annual season of persistent rains over the South Asian subcontinent associated with the northward migration of the Intertropical Convergence Zone (ITCZ). The India Meteorological Department (IMD) monitors and regularly publishes the onset, demise, and length of the rainy season (LRS) and provides a comparison with the climatological progress and demise every year (Extended Data Fig.\ref{fig: imd_isochrones_2020}). Despite recent advances\citep{gill1980some, gadgil2003indian, goswami2017dynamics, rajesh2020four, borah2020indian}, accurate definitions of characteristics of IM, such as onset, demise, and LRS, remain elusive. Here, we provide an objective and easily implementable definition for monitoring these characteristics by overcoming the identification of bogus onsets, a shortcoming of current approaches.  

The current definitions of IM are based on the persistence of local daily rainfall or regional weather monitoring of parameters such as winds or outgoing longwave radiation\citep{bombardi2020detection}. However, these definitions do not sufficiently identify if the local or regional precipitation is due to IM and if persistent rains will occur. Traditionally, MoK, the monsoon onset over Kerala (the state at the tip of the peninsula), is considered the first regional-scale onset of monsoon over the Indian peninsula. Subsequent local onsets are expected to occur following the northward migration of ITCZ associated with the first pulse of monsoon intraseasonal oscillation (MISO)\citep{goswami2005enso, Goswami2005}. MoK is often used as a benchmark for verifying onset forecasts. However, contrary to conventional wisdom, recent studies\citep{saha2023present, das2024dynamics} provide compelling evidence and dynamical arguments that the regional-scale onset over Northeast India (NEI) occurs before MoK in the middle of May. The NEI onset is driven either by the intrusion of extratropical planetary-scale Rossby waves or by quasi-biweekly monsoon oscillations\citep{das2024dynamics}.

A true large-scale onset and withdrawal occurs when the local weather phenomena become part of the planetary-scale monsoon system. Therefore, dynamically consistent true onsets of IM are those local onsets that are interconnected on a planetary-scale. Thus, we must be able to separate \textit{large-scale onsets (demises) from the local onsets (demises) at each location}. Unfortunately, for more than a century, the progress of IM, by necessity, has been defined by local onsets. However, local onsets defined only based on daily rainfall\citep{pai2020normal, ananthakrishnan1988onset, ananthakrishnan1991onset, wang2002rainy, joseph2006summer, wang2009objective, misra2018local, pai2009summer} are susceptible to false onsets and high uncertainties due to pre-monsoon showers and wind anomalies caused by transient weather systems\citep{flatau2001dynamics, noska2016characterizing} (Supplementary Information Sec. 1). We address this necessity by analysing the variations in spatial patterns of local onsets of precipitation using complex networks\citep{barabasi2016networks}. We isolate the clusters of local onsets that percolate in space and time to form a large-scale cluster of local onsets. To do so, we construct time-varying spatial-proximity complex networks (Methodology). A network is a set of nodes connected by links. Complex networks have been used extensively in recent years to study the Earth's climate and are often referred to as \textit{climate networks}\citep{dijkstra2019networks} (Supplementary Information Sec. 2). 

In our approach, at every time step, we construct spatial-proximity networks by connecting the geographical locations (nodes) that have undergone local onset and are spatial neighbours. Such networks have previously revealed the emergence of global structures from local connectivity in flows in engineering systems\citep{krishnan2019emergence, chopra2024evolution}. Here, we discover that the formation of the planetary-scale cluster of local onsets associated with the Indian monsoon occurs via two abrupt phase transitions accompanied by continuous phase growths. In a complex network, an abrupt transition occurs when two relatively small clusters merge to form a large cluster\citep{achlioptas2009explosive} (Supplementary Information Sec. 3). In our complex climate networks, we refer to these small and large clusters as the synoptic-scale and planetary-scale clusters, respectively. We discover that the first IM onset is associated with the first abrupt transition observed in our network. 

We define large-scale local onsets and demises of IM from the spatiotemporal evolution of the clusters of local onsets\citep{misra2018local} (Methodology, Extended Data Fig.\ref{fig: schematic}). We declare large-scale onset at each location when it becomes part of the largest cluster post the first abrupt transition. This definition proves to be highly reliable, as the strength and northward progress of the monsoon are maintained following the second abrupt transition. Our approach thus integrates the regional onsets over NEI and MoK within a single framework and busts the myth that IM starts with the MoK. Further, our method is apt for studying anomalous years and the contributions of evolutionary differences in the interseasonal variability of monsoon. 

\section*{Results}
\begin{figure}[h!]
	\centering
 	\includegraphics[width=1\textwidth]{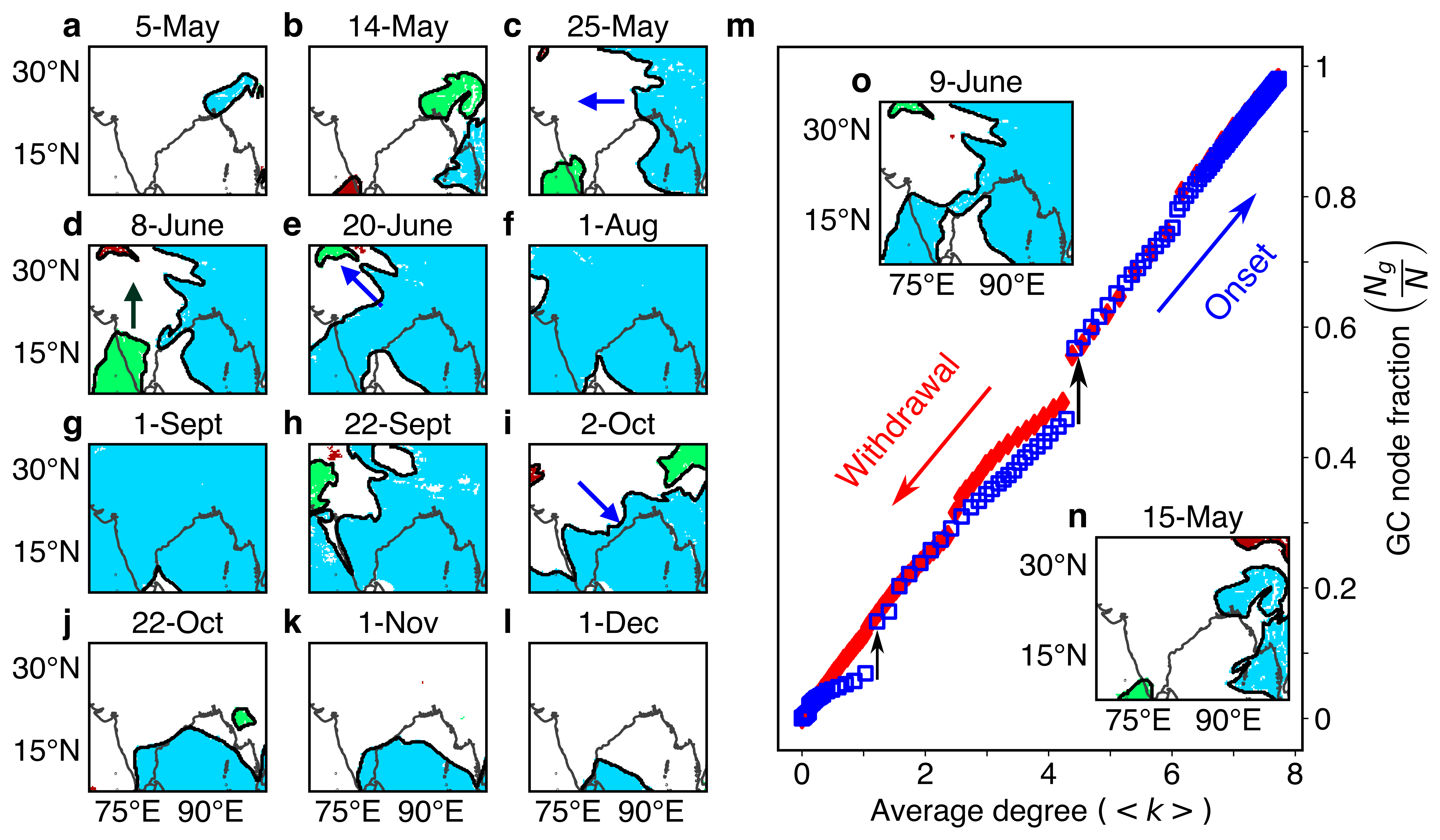}
	\caption{\doublespacing\textbf{Phase transitions in complex climate networks define the onset and withdrawal of the Indian Monsoon.} The spatiotemporal evolution of the connected components of the time-varying spatial-proximity networks is shown in (a) - (l). The giant connected component (GC) is in blue, followed by the second and third largest connected components in green and red, respectively. The connected components represent the clusters of the local onsets. (a) - (f) indicates the onset and advance of the monsoon over the Indian peninsula, followed by the retreat and withdrawal in (g) - (l). The network phase diagram (m) shows the variation of the fraction of locations in the largest cluster (GC node fraction) with the average number of links per location (average degree). The two abrupt transitions (black arrows) near $<\!k\!>\,\,=1$ and $<\!k\!>\,\,=4.5$ are associated with the merging of components over (n) NEI and over the (o) Indian peninsula, respectively, which are reminiscent of phase transitions. Such abrupt transitions are the fundamental characteristic of large-scale onsets, that occur over the NEI and peninsula on $15^{th}$ May and $9^{th}$ June, respectively. A gradual growth of the giant component occurs after the two components over NEI and peninsula merge, depicting a continuous phase transition over Northwest India. An animation is available online.} 
	\label{fig: components}
\end{figure}

With a rich history of monsoon research, IMD established the progress and retreat of IM from normal onsets and demise first\citep{india1943climatological} in 1943 based on rainfall data between 1901 and 1940. These progression curves were updated using more recent and reliable rainfall data\citep{pai2020normal}. However, these isochrones are based on local onset and demise. In the absence of a method to separate the large and local-scale onset (and demises), these isochrones served as the only reference for almost eight decades. Here, using complex network analysis, we disentangle the large-scale onsets (demises) from the local-scale onsets (demises) and present a novel view of the progress and retreat of IM.

\subsection*{True climatological IM onsets and demises as large-scale local onsets and demise}
The progress and retreat of IM are evident when we track the evolution of the clusters of interconnected climatological local onsets and demise between $1^{st}$ April and $31^{st}$ December (summarised in Fig.\ref{fig: components}). Notable are those mesoscale clusters that first appeared in Northeast India (NEI) in late April and coalesced into a synoptic-scale cluster. Subsequently, this cluster merges with another synoptic-scale cluster moving northward from the southeastern region of the Bay of Bengal on $15^{th}$ May. This interaction represents the first abrupt phase transition in size of largest cluster of the network (Fig.\ref{fig: components}m), marking the formation of a planetary-scale cluster over NEI (Fig.\ref{fig: components}n) that expands westwards. We define large-scale onset date at a specific location when this location becomes part of the planetary-scale cluster formed after the first abrupt transition.
 
Meanwhile, small clusters start appearing over the southwestern peninsula and the Arabian Sea from early May but remain constrained to a synoptic-scale. The second abrupt phase transition occurs on $9^{th}$ June (Fig.\ref{fig: components}m) when the synoptic-scale southern and planetary-scale NEI clusters merge to form a larger planetary-scale structure (Fig.\ref{fig: components}o). We refer to this event as the large-scale onset over the peninsula instead of traditional MoK. Following this transition, a combined front of the large-scale cluster of local onsets expands rather slowly towards the north and northwest, and covers the whole country by August. The retreat from Northwest India starts around $22^{nd}$ September and rapidly retreats across most of the country by $22^{nd}$ October.

\subsection*{Progress and Withdrawal of Indian Monsoon (IM): Large-scale versus Local}
\begin{figure}[h!]
    \centering
    \includegraphics[width=1\textwidth]{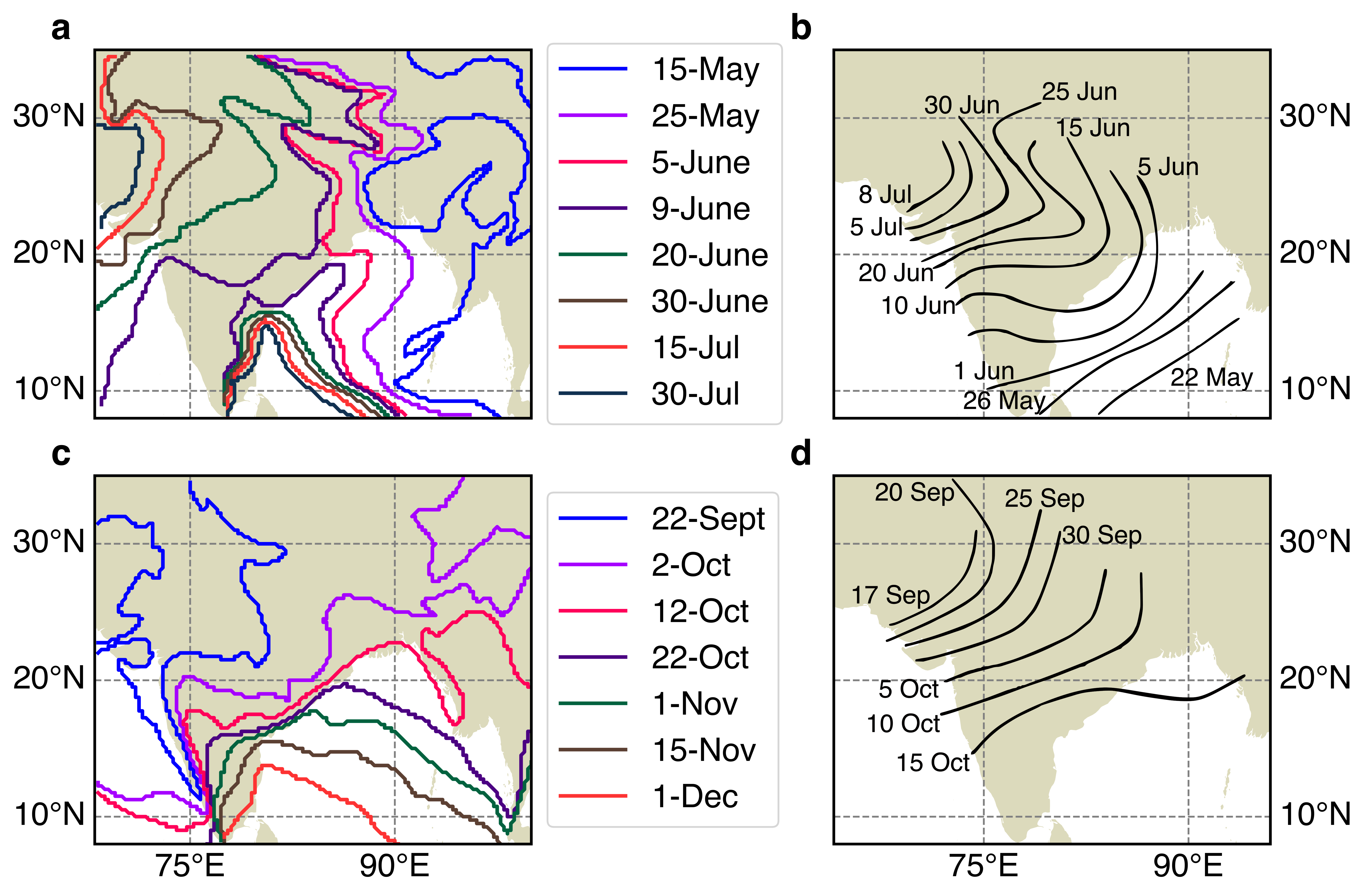}
    \caption{\setstretch{1.48}\textbf{Our climatological onset and withdrawal dates are fundamentally different from those given by the India Meteorological Department.} (a) Large-scale onset and (c) large-scale withdrawal of the Indian monsoon constructed from time-varying spatial-proximity networks. Also shown for comparison are the isochrones reproduced from \citet{pai2020normal} using modified IMD’s criteria. The monsoon advances westwards after NEI onset on $15^{th}$ May (blue curve in (a)), followed by a northwest progression post the MoK on $9^{th}$ June (purple curve in (a)). This westward expanding NEI component merges with the synoptic-scale component over the southwestern peninsula and is vitally important for the occurrence of MoK. The onset isochrones given by IMD in (b) lack information about the NEI onset, while our network analysis captures the NEI onset and MoK in a unified framework. The monsoon withdraws along the southeast direction, and the southeastern tip of the Indian peninsula undergoes large-scale withdrawal in December (orange curve in (c)). In contrast, the IMD does not give any information on the withdrawal of monsoon (d) post $15^{th}$ October. However, this withdrawal information is inherently highlighted by the network analysis. We also show that our findings are robust by comparing the isochrones obtained from ERA5 reanalysis precipitation data in (a) and (c) with that of the IMDAA reanalysis and IMERG satellite data (Extended Data Fig.\ref{fig: isochrones_all_datasets}). The main features of IM are faithfully reproduced using all three daily rainfall datasets.}
    \label{fig: isochrones}
\end{figure}

We construct the isochrones of the large-scale climatological advance and retreat of IM at 10-day intervals and compare them with those of IMD\citep{pai2020normal} primarily based on local onsets (Fig.\ref{fig: isochrones}). A major difference between our large-scale onsets and the IMD onsets is that the large-scale onsets over NEI occur prior to those over the peninsula. Interestingly, our climatological IM onset over NEI on $15^{th}$ May aligns perfectly with another independent objective method that involves identifying a low-level cyclonic vortex\citep{das2024dynamics} on $18^{th}$ May. The southeastern peninsula, being under the rain shadow of large-scale monsoon winds, gets little rain during the summer monsoon, while it gets most of the rain in winter during retreating monsoon\citep{sanap2019heavy}. Our method naturally avoids this rain shadow region during the summer monsoon and, therefore, supports the objectivity of our method. Our physical and dynamically consistent definition of large-scale onset and evolution of IM unravels that the true large-scale onset of IM first occurs over NEI, and expands westwards covering the country. This is a paradigm shift from the prevailing notion over a century that the first IM onset happens in southern peninsular India (also known as MoK) and covers the country by northward and westward expansion.

\subsection*{Monsoon Intra-Seasonal Oscillations (MISO): Key to defining phase transitions and large-scale onsets}
A large-scale inhibition restricts the northward propagation of ITCZ and MISO to the south\citep{goswami2005enso, lau2012MISO} of $10^{\circ}$N. A symmetric instability\citep{tomas1997role, krishnakumar1998possible} is required to overcome inhibition and invigorate the ITCZ, and to establish conditions for the feedback\citep{jiang2004structures} that initiates the first northward propagating pulse of MISO\citep{goswami2005enso, lau2012MISO}. In May and early June, there is often a MISO pulse\citep{lau2012MISO} that propagates approximately to, but not beyond, $10^{\circ}$N (Fig.2.15). This leads to the identification of a bogus onset based on regional onsets in Kerala (MoK). The next MISO pulse takes about another 20 days (half period of MISO) to arrive. The key to a successful large-scale MoK definition is to be able to tag MoK only to the first northward propagating MISO pulse\citep{goswami2005enso, wang2006asian}.
\begin{figure}[h!]
    \centering
    \includegraphics[width=0.97\textwidth]{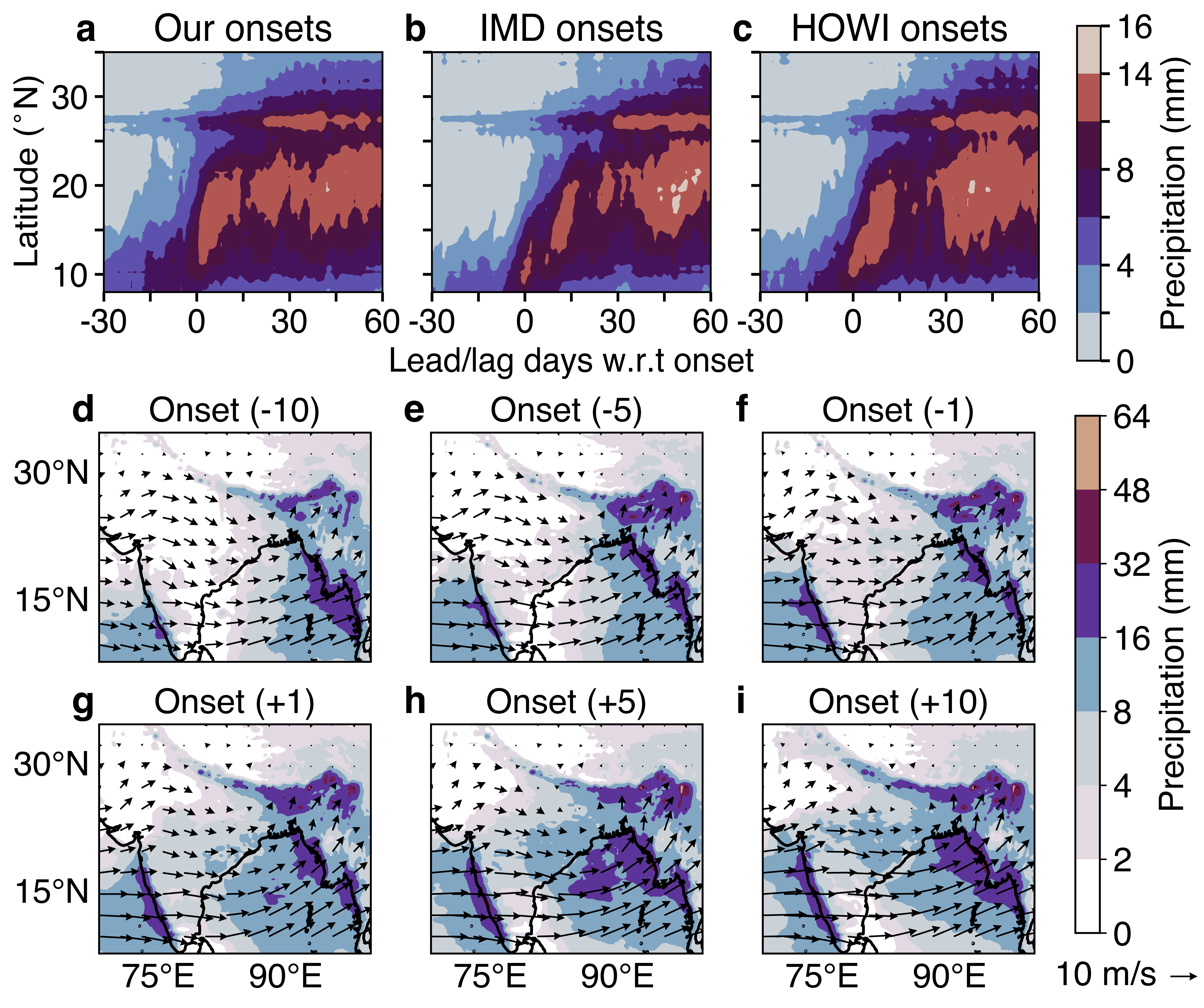}
    \caption{\textbf{Our network-based onsets show a rapid and consistent northward propagation of rainfall when compared with the existing definitions\citep{pai2020normal, fasullo2003hydrological}.} Hovm\"oller diagram for the zonally averaged precipitation between $70^\circ$E - $90^\circ$E from 30 days before to 60 days after the (a) large-scale onset determined from our network, (b) IMD onsets and (c) Hydrological Onset and Withdrawal index (HOWI)\citep{fasullo2003hydrological} onsets. The precipitation increases abruptly in (a) from nearly $6$ mm/day before the onset to more than $12$ mm/day in 2-3 days of the large-scale onset. This indicates that the rainfall occurs in a northward propagating band (ITCZ), moving at a rate of approximately $1^{\circ}$N/day and then establishes at $20^{\circ}$N. This is the average rate of northward propagation of MISO\citep{Goswami2005}. The IMD onsets in (b) show that the rain band does not propagate beyond $10^{\circ}$N after MoK, indicating a preponderance in the detection of bogus onsets in the operational MoK. The HOWI onsets in (c) and Extended Data Fig.\ref{fig: HOWI_MoK_composites} are associated with the first pulse of northward propagating MISO in most years. However, this index cannot be used to obtain the progression of IM since it is a regional-scale definition based on averaging vertically integrated moisture transport over a large area. (d)-(i) show the lead-lag composites of wind and precipitation from 1940-2023 centred w.r.t the network-based onset dates over the 14 stations in Kerala. IMD uses the same 14 locations to declare the MoK. The temporal average of wind and precipitation at (d) 10 days, (e) 5 days, (f) 1 day before the MoK and (g) 1 day, (h) 5 days, (i) 10 days after the MoK. The onset is associated with the strengthening of large-scale low-level winds over the entire Indian monsoon region, leading to significant organised rainfall on the windward side of both the Western Ghats and the Myanmar orography.}
    \label{fig: large_composites}
\end{figure}

The lead-lag composites (Fig.\ref{fig: large_composites}a-f) of precipitation and 850 hPa winds for large-scale MoK detected by our approach show that the precipitation and winds organise at a large-scale over the Indian monsoon region post the large-scale onset. Furthermore, the zonally averaged large-scale precipitation between $70^\circ$E - $90^\circ$E as a function of latitude and time (Fig.\ref{fig: large_composites}g) shows a sudden phase transition. The precipitation increases twofold, from 6 to 12 mm/day, and remains sustained post the occurrence of large-scale onset over the peninsula. Most importantly, every large-scale onset over the peninsula identifies the first completely northward propagating MISO, showing a robust northward propagation of rainband to $20^{\circ}$N in 10 days. This indicates that there is no identification of bogus onsets. 

In contrast, the lead-lag composites with respect to MoK obtained by IMD onsets (Extended Data Fig.\ref{fig: IMD_composites}) and local onsets (Extended Data Fig.\ref{fig: local_composites}) do not delineate the large-scale monsoon winds and lack organised precipitation, indicating the dominance of synoptic activity. Furthermore, the lead-lag composites around MoK defined by the tropospheric temperature gradient method\citep{xavier2007objective} (Extended Data Fig.\ref{fig: TT_MoK_composites}) and that based on vertically integrated moisture flux using change point index\citep{walker2016onset} (Extended Data Fig.\ref{fig: CHP_MoK_composites}) show that the first pulse of northward propagating MISO occurs about 20 days after the identified onsets, indicating a considerable probability of detection of bogus onsets. 

We also construct similar lead-lag composites with respect to the large-scale onsets over NEI ($22^{\circ}$N-$26^{\circ}$N, $90^{\circ}$E-$94^{\circ}$E). These composites (Extended Data Fig.\ref{fig: NEI_composites}) show organised wind and precipitation over NEI, driven by a low-level cyclonic vortex as proposed by \citet{das2024dynamics}

\subsection*{Length of the rainy season (LRS) from large-scale onset and demise}
\begin{figure}[h!]
    \centering
    \includegraphics[width=1\textwidth]{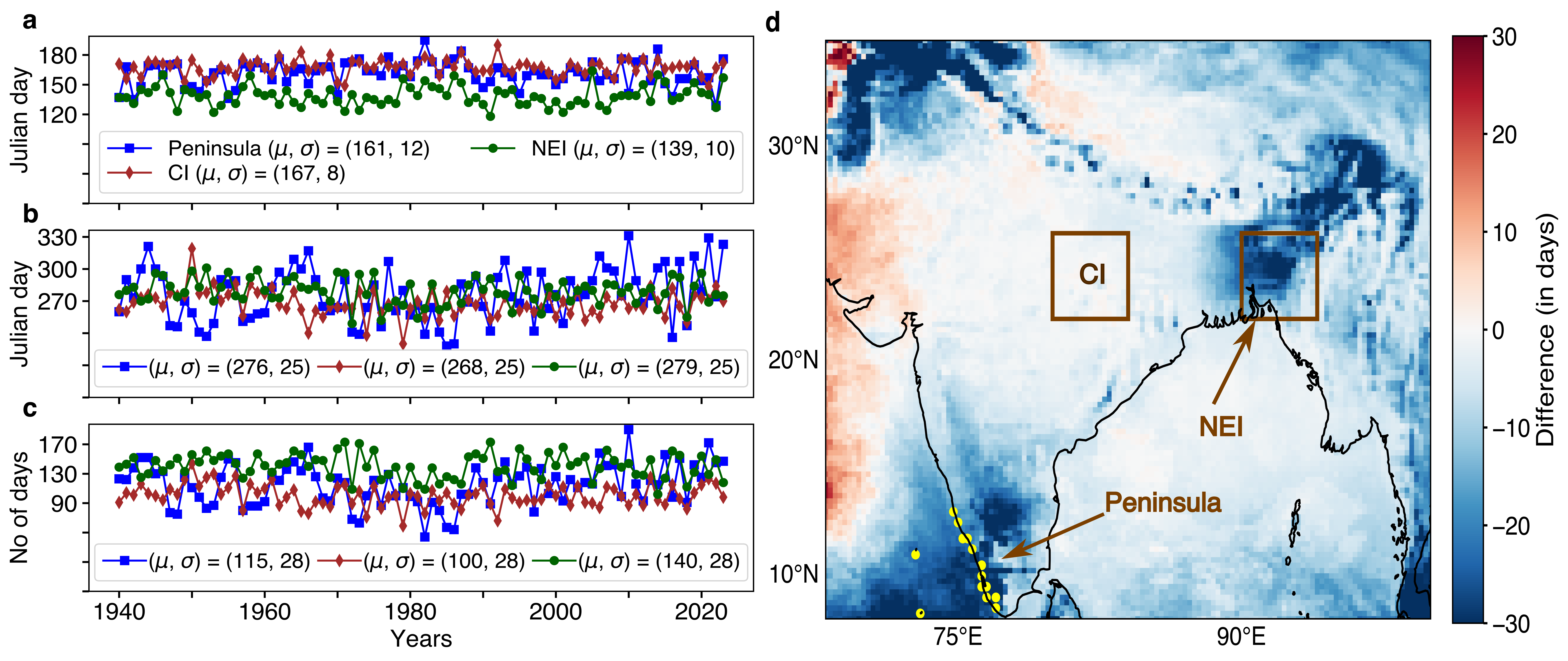}
    \caption{\doublespacing\textbf{Interannual variability of the characteristics of Indian Monsoon from 1940-2023.} Time series of (a) large-scale onset, (b) withdrawal, and (c) length of the rainy season (LRS) over three different regions: the Indian peninsula (14 rain gauge stations in Kerala), CI, and NEI are shown in blue, brown, and green respectively. The large-scale onset and withdrawal, along with their mean ($\mu$), are expressed in terms of Julian days in (a) and (b), while the large-scale length of the rainy season with its mean ($\mu$) is expressed in number of days in (c). The standard deviation ($\sigma$) for (a), (b), and (c) are expressed in number of days. The mean LRS over the Indian peninsula of 115 days in (c) is close to the conventional definition of 122 days of the Indian summer monsoon season (June-September). However, the season defined by our approach is shifted by about 7 days, thus challenging the credence of the conventional definition. The large-scale LRS over the NEI of 140 days determined from our method is consistent with the tropospheric temperature method\citep{xavier2007objective}, which is 144 days. (d) shows the difference between large and local-scale LRS. The large-scale LRS is $15-20$ days shorter than the local-scale LRS over the southwestern peninsula and NEI. The 14 stations in Kerala are marked in yellow, and the CI and NEI regions are enclosed within brown boxes.}
    \label{fig: interannual_var}
\end{figure}

Since the quantum of monsoon rainfall is determined by LRS, any error in it has direct implications in policy-making for water and food security. We argue that the LRS defined by large-scale onset and demise is more reliable than that defined based on local onset and demise. Figure \ref{fig: interannual_var}d shows that the large-scale LRS is nearly $15-20$ days shorter than the local-scale LRS in the southwestern peninsula and NEI. This is because the synoptic-scale onset precedes the large-scale onset over the southwestern peninsula. In the case of NEI, vigorous mesoscale activity in April obfuscates local onsets, whereas large-scale organisation occurs only in the middle of May. In contrast, large-scale LRS is approximately $10-15$ days longer than local-scale LRS over northwest India. Local onsets are tuned to vigorous synoptic activity. In semi-arid regions, this activity occurs only after large-scale organisation; therefore, local onsets occur after large-scale onsets. 

The accumulated rainfall during IM depends partly on the LRS and the subseasonal rainfall activity. Therefore, significant differences in large-scale and local-scale LRS may change the perceived relationship between LRS and seasonal rainfall. We define LRS rainfall as the daily rainfall accumulated for the LRS days and find correlations between large-scale LRS rainfall and large-scale onset dates, demise dates, LRS and June-September rainfall based on data between 1940 and 2023 (Extended Data Fig.\ref{fig: LRS_prep_correlation}). As may be expected, large-scale LRS and demise dates correlate positively with LRS rainfall in the whole region, while onset dates correlate negatively with it. We note that LRS rainfall and rainfall from June to September have a strong positive correlation over most of the country (Extended Data Fig.\ref{fig: LRS_prep_correlation}d).

\section*{Discussion}
The IM being a planetary-scale phenomenon, the onset, demise, and LRS at different locations must be interconnected on a planetary-scale. Yet, all definitions of onset, demise, and LRS, along with associated progress and retreat of IM so far, are based on local onset and demise. Hence, they are prone to detection of bogus onset and demise, making the information practically inutile to the users. Even after a century of research on IM, identifying the large-scale onset and demise at each location is lacking. Here, using time-varying spatial-proximity networks, we describe a dynamically consistent method to separate large-scale from local onset, demise and LRS for the first time. We unravel that the first large-scale IM onset occurs over the NEI and covers the country by expanding westward. Our discovery is a paradigm shift from the century old notion based on local onsets that the first IM onset happens over the southern peninsular India and covers the country by northward and westward expansion.

Four distinct spatial and temporal scales, namely cumulonimbus ($\sim$1-10 km, hour), mesoscale convective clusters (MCS) ($\sim$100-500 km, day), superclusters or synoptic-scales ($\sim$1000-3000 km, week), and the Madden Julian Oscillations and Monsoon Intraseasonal Oscillations envelop on planetary-scale ($\sim$ 10000 km, months) illustrate the multiscale organisation of convection in tropics\citep{moncrieff2010multiscale, moncrieff2012multiscale}. The physical basis for using time-varying spatial-proximity networks comes from the fact that the spatial pattern and size of interconnected clusters of local onsets are results of the organisation of daily rainfall by meteorological phenomena with different characteristic spatial scales. Our complex network analysis indeed finds that the clustering of interconnected local onsets falls into these categories: (i) small meso-scale clusters, (ii) large synoptic-scale clusters, and (iii) giant super-synoptic clusters.

We discover that the progression and withdrawal of IM are associated with two discontinuous phase transitions interspersed with continuous phase transitions (Fig.\ref{fig: components}). The first phase transition is marked by a large-scale onset over the NEI, thereby addressing the fundamental fact that the NEI onset occurs before MoK. The monsoon onset over NEI expands westward (Fig.\ref{fig: isochrones}) and is essential for the second phase transition associated with large-scale onset over the peninsula, including MoK. The second phase transition results in the formation of a cluster of local onsets interconnected at a planetary-scale. Physically, this transition represents the establishment of thermodynamic conditions appropriate for the northward progression of the ITCZ and the IM. Subsequently, we provide an objective definition of dates of large-scale monsoon onset and withdrawal based on the dynamics and transitions in the network. Farmers and policymakers demand local information on the onset, demise, and LRS for agricultural planning and water resource management. \textbf{Unlike existing large-scale definitions \citep{das2024dynamics, pai2020normal, fasullo2003hydrological, xavier2007objective}, our definition has monitoring capabilities and provides true large-scale monsoon onset dates, withdrawal dates, and LRS at any desired location.} Our findings represent a significant advance on this objective and have been a major area of research due to the keen interest of the user community in predicting bogus free MoK. Moreover, the simplicity and robustness of our method make it applicable to other monsoon systems in the world.

Our finding that the NEI onset is critical for the annual cycle of evolution of IM rainfall brings into focus the potential role of extratropical systems\citep{das2024dynamics} in driving the interannual variability and complexity of IM. Therefore, we need to revise the prevailing notion that it is the ITCZ that drives IM and its variability. We believe our findings also bring out new insight into how tropics and extratropics could interact. Furthermore, significant differences are expected between the variability of LRS rainfall and fixed-season June to September (JJAS). This means that potential predictability studies of seasonal JJAS mean IM rainfall thus far warrant re-examination.  

\newpage
\section*{Materials and Method}

\subsection*{Data}

In this study, we use precipitation data from two reanalysis datasets: the fifth-generation European Centre for Medium-Range Weather Forecasts (ECMWF) reanalysis, or ERA5\citep{hersbach2020era5} and the Indian Monsoon Data Assimilation and Analysis, or IMDAA\citep{rani2021imdaa} from the National Centre for Medium-Range Weather Forecasts (NCMWRF). Further, we use precipitation data from the Integrated Multi-Satellite Retrievals for Global Precipitation Mission (GPM IMERG Final V07) satellite dataset given by NASA\citep{huffman2023gpm}. The spatial domain considered in this study is restricted to the geographical region from 8$^{\circ}$N-35$^{\circ}$N and 68$^{\circ}$E-100$^{\circ}$E. The raw precipitation data is then preprocessed to obtain an appropriate spatial and temporal resolution. The spatial resolution of the precipitation data for ERA5 and IMDAA datasets is approximately $0.25^\circ \times 0.25^\circ$ while that for the IMERG dataset is $0.2^\circ \times 0.2^\circ$. The temporal resolution of the precipitation data is 1 day.  

\subsection*{Methodology}

The steps followed for network construction and analysis of the spatiotemporal evolution of the connected components (Extended Data Fig.\ref{fig: schematic}) are as follows:
\begin{enumerate}
    \item \textbf{Determination of local onset and withdrawal}: We find the local onset and withdrawal using the criteria of \citet{misra2018local}, where the minima and maxima of the daily cumulative precipitation anomaly curve represent the onset and withdrawal at any given location, respectively.
    \begin{equation}
        CPA(lat, \, lon, \, dy, \, yr) = \sum_{dy \, = \, 1 \, Jan}^{dy}\left[P(lat, \, lon, \, dy, \, yr) - \overline{\overline{C}}(lat, \, lon)\right]
        \label{eq: misra_criteria}
    \end{equation}
    where \textit{P} is the precipitation, and CPA is the daily cumulative precipitation anomaly. The terms \textit{lat, lon, dy}, and \textit{yr} denote the latitude, longitude, day, and year respectively. $\overline{\overline{C}}$ is the annual mean climatology at a particular location, given by
    \begin{equation}
        \overline{\overline{C}}(lat, \, lon) = \dfrac{\sum_{yr \, = \, 1940}^{yr \, = \, 2023}\left(\sum_{dy \, = \,1 \,Jan}^{dy \, = \, 31 \, Dec} P(lat, \, lon, \, dy, \, yr)\right)}{N_{yrs}\times N_{days}}
        \label{eq: annual_mean_clim}
    \end{equation}
    where $N_{yrs}$ and $N_{days}$ correspond to 84 years with 365 (or 366 for leap) years, respectively. The onset day ($t_{onset}$), withdrawal day ($t_{withdrawal}$), and length of the rainy season (LRS) at that location are given by
    \begin{align}
        \begin{gathered}
            t_{onset}(lat, \, lon, \, yr) = argmin\left[CPA(lat, \, lon, \, dy, \, yr)\right] + 1 \\
            t_{withdrawal}(lat, \, lon, \, yr) = argmax\left[CPA(lat, \, lon, \, dy, \, yr)\right] + 1 \\
            LRS(lat, \, lon, \, yr) = t_{withdrawal}(lat, \, lon, \, yr) - t_{onset}(lat, \, lon, \, yr)
        \end{gathered}
        \label{eq: ons_with_def}
    \end{align}

    We then obtain the climatological or normal local onsets (and demises) by averaging the local onsets (and demises) from 1940-2023.

    \item \textbf{Construction of time-varying spatial-proximity networks}: We construct time-varying spatial-proximity networks for each day from $1^{st}$ April to $31^{st}$ December for the past 84 years, from 1940 to 2023. The geographical locations (or the grid point) in the spatial domain represent the nodes, and the links are established based on the satisfaction of the two criteria:
    
    C$_{1}$. At any time instant \textit{t} of a given year, \textit{t} lies between the local onset and local withdrawal day:
    \begin{align}
        \begin{gathered}
            t_{withdrawal}(i) > t(i) > t_{onset}(i) \\
            t_{withdrawal}(j) > t(j) > t_{onset}(j)
        \end{gathered}
        \label{eq: C1}
    \end{align}
    C$_{2}$. Both nodes are in spatial-proximity:
    \begin{equation}
        N_{sp} = \left\{(i, \, j); \, \Delta lat(i, \, j) \,\cap \,\Delta lon(i, \, j) \leq d\right\}
        \label{eq: C2}
    \end{equation}
    where \textit{i, j} represents a particular node, and \textit{d} is the distance between the latitudes and longitudes of both nodes. Here, we choose $d = 1$, i.e., we consider the Moore neighbourhood, excluding the central cell itself. A Moore neighbourhood of a 2D square lattice consists of a central cell and eight surrounding cells. We connect only the neighbouring locations as we aim to study how the local interactions between the nodes lead to the emergence over a planetary-scale. The emergence here refers to the large-scale connectivity of locations that have undergone local onsets. Enforcing the condition of spatial-proximity in the network allows us to study how the local onset and withdrawal of monsoon are associated with the large-scale onset and withdrawal.
    The adjacency matrix is then constructed using the criteria C$_1$ and C$_2$:
    \begin{equation}
        A_{ij} =
        \begin{cases}
            1, & \text{if C$_{1}$ and C$_{2}$ are satisfied} \\ 
            0, & \text{otherwise}
        \end{cases}
    \end{equation}
    \label{eq: adj_mat}
    We remove the links only when a node undergoes local withdrawal. This helps us capture the progression of monsoon throughout the Indian peninsula from onset to withdrawal. We also construct the time-varying spatial-proximity networks for the climatological local onsets (and demise).
    
    \item \textbf{Analysis of the connected components of the network}: As the Indian monsoon is a planetary-scale phenomenon, its local onsets must be connected on the planetary-scale. We analyse the spatiotemporal evolution of the connected components of the network and define the onset of the Indian monsoon only when the local onsets are connected on the scale of the largest connected component, called the giant component or largest cluster. We use the information of the largest cluster because, initially, during the pre-monsoon season, there will be small clusters of local onsets, which eventually merge to form a large cluster or giant component during the onset of monsoon. This large cluster then grows in size as the monsoon advances until it reaches its peak. Once the monsoon attains its peak, the largest cluster then shrinks in size and disintegrates into small fragments during the withdrawal phase.

    \item \textbf{Large-scale definition of onset and withdrawal of Indian Monsoon}: We define the large-scale onset of monsoon at a given node based on the following two criteria:
    
    C$_3$. At any time instant \textit{T} of a given year, \textit{T} lies after the day when the first abrupt transition occurs in the network phase diagram (Fig.\ref{fig: components}m, n), which corresponds to the large-scale onset over NEI.
    \begin{equation}
        T_{onset} > T_{abrupt^{1}}
        \label{eq: large_ons_C1}
    \end{equation}
    C$_4$. The node at time \textit{T$_{onset}$} must be a part of the giant component (GC).
    \begin{equation}
        N_{T_{onset}} = \left\{(i, \, j); \, (i, \, j) \, \in \, (GC)_{T_{onset}}\right\} 
        \label{eq: large_ons_C2}
    \end{equation}
    A particular node is said to undergo large-scale onset if both criteria C$_3$ and C$_4$ are satisfied. 
    
    The first day when the node N$_{T}$ is no longer part of the GC is declared to be the large-scale withdrawal date for that particular node. Mathematically, the large-scale withdrawal criteria at any location are given by Eq. \ref{eq: large_with_C3}-\ref{eq: large_with_C4}.
    
    \begin{equation}
        N_{T_{withdrawal}} = \left\{(i, \, j); \, (i, \, j) \, \notin \, (GC)_{T_{withdrawal}}\right\} 
        \label{eq: large_with_C3}
    \end{equation}
    \begin{equation}
        N_{T_{withdrawal} - 1} = \left\{(i, \, j); \, (i, \, j) \, \in \, (GC)_{T_{withdrawal} - 1}\right\} 
        \label{eq: large_with_C4}
    \end{equation}
    
    We note that the magnitude of the jump in the first abrupt transition over NEI varies throughout the years. Therefore, we have chosen a threshold of 0.04 as the jump in the node fraction of the largest cluster in the network phase diagram. This enables us to capture the abrupt transition over NEI for all the years. This jump of 0.04 corresponds to an area equivalent to $5^{\circ} \times 5^{\circ}$ geographical region. However, the spatiotemporal evolution of the connected components for all the years remains qualitatively similar to the evolution of the connected components presented in the Results section (Progress and Withdrawal of IM: Large-scale versus Local) discussed above.
\end{enumerate}

\subsection*{Data availability}
The three precipitation datasets that support the findings of this study are publicly available online:
\begin{enumerate}
    \item ERA-5 Global Reanalysis: \href{https://doi.org/10.24381/cds.adbb2d47}{https://doi.org/10.24381/cds.adbb2d47}
    \item IMDAA Reanalysis: \href{https://doi.org/10.1175/JCLI-D-20-0412.1}{https://doi.org/10.1175/JCLI-D-20-0412.1}
    \item GPM IMERG V07 satellite data:
    \href{https://doi.org/10.5067/GPM/IMERGDF/DAY/07}{https://doi.org/10.5067/GPM/IMERGDF/DAY/07}
\end{enumerate}

\subsection*{Code availability}
The codes used in this study can be made available upon reasonable request from the corresponding author.



\bibliography{ref.bib}

\begin{thebibliography}{42}
\providecommand{\natexlab}[1]{#1}
\providecommand{\url}[1]{\texttt{#1}}
\expandafter\ifx\csname urlstyle\endcsname\relax
  \providecommand{\doi}[1]{doi: #1}\else
  \providecommand{\doi}{doi: \begingroup \urlstyle{rm}\Url}\fi

\bibitem[Achlioptas et~al.(2009)Achlioptas, D'Souza, and
  Spencer]{achlioptas2009explosive}
D.~Achlioptas, R.~M. D'Souza, and J.~Spencer.
\newblock Explosive percolation in random networks.
\newblock \emph{Science}, 323\penalty0 (5920):\penalty0 1453--1455, 2009.

\bibitem[Ananthakrishnan and Soman(1988)]{ananthakrishnan1988onset}
R.~Ananthakrishnan and M.~Soman.
\newblock The onset of the southwest monsoon over kerala: 1901--1980.
\newblock \emph{Int. J. Climatol}, 8\penalty0 (3):\penalty0 283--296, 1988.

\bibitem[Ananthakrishnan and Soman(1991)]{ananthakrishnan1991onset}
R.~Ananthakrishnan and M.~Soman.
\newblock The onset of the southwest monsoon in 1990.
\newblock \emph{Curr. Sci.}, 61:\penalty0 447--453, 1991.

\bibitem[Barabási and Pósfai(2016)]{barabasi2016networks}
A.~Barabási and M.~Pósfai.
\newblock \emph{Network Science}.
\newblock Cambridge University Press, Cambridge, 2016.

\bibitem[Bombardi et~al.(2020)Bombardi, Moron, and
  Goodnight]{bombardi2020detection}
R.~J. Bombardi, V.~Moron, and J.~S. Goodnight.
\newblock Detection, variability, and predictability of monsoon onset and
  withdrawal dates: A review.
\newblock \emph{Int. J. Climatol}, 40:\penalty0 641--667, 2020.

\bibitem[Borah et~al.(2020)Borah, Venugopal, Sukhatme, Muddebihal, and
  Goswami]{borah2020indian}
P.~J. Borah, V.~Venugopal, J.~Sukhatme, P.~Muddebihal, and B.~Goswami.
\newblock Indian monsoon derailed by a north atlantic wavetrain.
\newblock \emph{Science}, 370\penalty0 (6522):\penalty0 1335--1338, 2020.

\bibitem[Chopra et~al.(2024)Chopra, Mittal, and Sujith]{chopra2024evolution}
G.~Chopra, S.~Mittal, and R.~Sujith.
\newblock Evolution of clusters of turbulent reattachment due to shear layer
  instability in flow past a circular cylinder.
\newblock \emph{Phys. Fluids}, 36\penalty0 (1), 2024.

\bibitem[Das et~al.(2024)Das, Goswami, Mahanta, Saha, and
  Goswami]{das2024dynamics}
S.~Das, D.~J. Goswami, R.~Mahanta, P.~Saha, and B.~Goswami.
\newblock Dynamics of may ‘onset’of indian summer monsoon over northeast
  india.
\newblock \emph{Quarterly Journal of the Royal Meteorological Society},
  150\penalty0 (764):\penalty0 4533--4549, 2024.

\bibitem[Dijkstra et~al.(2019)Dijkstra, Hern{\'a}ndez-Garc{\'\i}a, Masoller,
  and Barreiro]{dijkstra2019networks}
H.~A. Dijkstra, E.~Hern{\'a}ndez-Garc{\'\i}a, C.~Masoller, and M.~Barreiro.
\newblock \emph{Networks in Climate}.
\newblock Cambridge University Press, Cambridge, 2019.

\bibitem[Fasullo and Webster(2003)]{fasullo2003hydrological}
J.~Fasullo and P.~Webster.
\newblock A hydrological definition of indian monsoon onset and withdrawal.
\newblock \emph{J. Clim.}, 16\penalty0 (19):\penalty0 3200--3211, 2003.

\bibitem[Flatau et~al.(2001)Flatau, Flatau, and Rudnick]{flatau2001dynamics}
M.~K. Flatau, P.~J. Flatau, and D.~Rudnick.
\newblock The dynamics of double monsoon onsets.
\newblock \emph{J. Clim.}, 14\penalty0 (21):\penalty0 4130--4146, 2001.

\bibitem[Gadgil(2003)]{gadgil2003indian}
S.~Gadgil.
\newblock The indian monsoon and its variability.
\newblock \emph{Annu. Rev. Earth Planet. Sci.}, 31\penalty0 (1):\penalty0
  429--467, 2003.

\bibitem[Gill(1980)]{gill1980some}
A.~E. Gill.
\newblock Some simple solutions for heat-induced tropical circulation.
\newblock \emph{Q. J. R. Meteorol. Soc}, 106\penalty0 (449):\penalty0 447--462,
  1980.

\bibitem[Goswami(2005)]{Goswami2005}
B.~N. Goswami.
\newblock \emph{South Asian monsoon}, chapter South Asian monsoon, pages
  19--61.
\newblock Springer Berlin Heidelberg, Berlin, Heidelberg, 2005.

\bibitem[Goswami and Chakravorty(2017)]{goswami2017dynamics}
B.~N. Goswami and S.~Chakravorty.
\newblock Oxford research encyclopedia of climate science.
\newblock In \emph{Dynamics of the Indian Summer Monsoon Climate}. Oxford
  University Press, Oxford, 2017.
\newblock \doi{10.1093/acrefore/9780190228620.013.613}.

\bibitem[Goswami and Xavier(2005)]{goswami2005enso}
B.~N. Goswami and P.~K. Xavier.
\newblock Enso control on the south asian monsoon through the length of the
  rainy season.
\newblock \emph{Geophys. Res. Lett.}, 32\penalty0 (18), 2005.

\bibitem[Hersbach et~al.(2020)Hersbach, Bell, Berrisford, Hirahara,
  Hor{\'a}nyi, Mu{\~n}oz-Sabater, Nicolas, Peubey, Radu, Schepers,
  et~al.]{hersbach2020era5}
H.~Hersbach, B.~Bell, P.~Berrisford, S.~Hirahara, A.~Hor{\'a}nyi,
  J.~Mu{\~n}oz-Sabater, J.~Nicolas, C.~Peubey, R.~Radu, D.~Schepers, et~al.
\newblock The era5 global reanalysis.
\newblock \emph{Q. J. R. Meteorol. Soc}, 146\penalty0 (730):\penalty0
  1999--2049, 2020.

\bibitem[Huffman et~al.(2023)Huffman, Stocker, Bolvin, Nelkin, and
  Jackson]{huffman2023gpm}
G.~Huffman, E.~Stocker, D.~Bolvin, E.~Nelkin, and T.~Jackson.
\newblock Gpm imerg final precipitation l3 1 day 0.1 degree x 0.1 degree v07,
  greenbelt, md, goddard earth sciences data and information services center
  (ges disc).
\newblock \emph{\\https://doi. org/10.5067/GPM/IMERGDF/DAY/07}, 2023.

\bibitem[{India Meteorological Department}(1943)]{india1943climatological}
{India Meteorological Department}.
\newblock \emph{Climatological Atlas for Airmen}.
\newblock India Meteorological Department, New Delhi, 1943.

\bibitem[Jiang et~al.(2004)Jiang, Li, and Wang]{jiang2004structures}
X.~Jiang, T.~Li, and B.~Wang.
\newblock Structures and mechanisms of the northward propagating boreal summer
  intraseasonal oscillation.
\newblock \emph{J. Clim.}, 17\penalty0 (5):\penalty0 1022--1039, 2004.

\bibitem[Joseph et~al.(2006)Joseph, Sooraj, and Rajan]{joseph2006summer}
P.~Joseph, K.~Sooraj, and C.~Rajan.
\newblock The summer monsoon onset process over south asia and an objective
  method for the date of monsoon onset over kerala.
\newblock \emph{Int. J. Climatol.}, 26\penalty0 (13):\penalty0 1871--1893,
  2006.

\bibitem[Kathayat et~al.(2017)Kathayat, Cheng, Sinha, Yi, Li, Zhang, Li, Ning,
  and Edwards]{kathayat2017indian}
G.~Kathayat, H.~Cheng, A.~Sinha, L.~Yi, X.~Li, H.~Zhang, H.~Li, Y.~Ning, and
  R.~L. Edwards.
\newblock The indian monsoon variability and civilization changes in the indian
  subcontinent.
\newblock \emph{Sci. Adv.}, 3\penalty0 (12):\penalty0 e1701296, 2017.

\bibitem[Krishnakumar and Lau(1998)]{krishnakumar1998possible}
V.~Krishnakumar and W.~K.~M. Lau.
\newblock Possible role of symmetric instability in the onset and abrupt
  transition of the asian monsoon.
\newblock \emph{Journal of the Meteorological Society of Japan. Ser. II},
  76\penalty0 (3):\penalty0 363--383, 1998.

\bibitem[Krishnan et~al.(2019)Krishnan, Sujith, Marwan, and
  Kurths]{krishnan2019emergence}
A.~Krishnan, R.~I. Sujith, N.~Marwan, and J.~Kurths.
\newblock On the emergence of large clusters of acoustic power sources at the
  onset of thermoacoustic instability in a turbulent combustor.
\newblock \emph{J. Fluid Mech.}, 874:\penalty0 455--482, 2019.

\bibitem[Lau et~al.(2012)Lau, Waliser, and Goswami]{lau2012MISO}
W.~K.~M. Lau, D.~E. Waliser, and B.~N. Goswami.
\newblock \emph{South Asian monsoon}, pages 21--72.
\newblock Springer Berlin Heidelberg, Berlin, Heidelberg, 2012.
\newblock ISBN 978-3-642-13914-7.
\newblock \doi{10.1007/978-3-642-13914-7_2}.
\newblock URL \url{https://doi.org/10.1007/978-3-642-13914-7_2}.

\bibitem[Misra et~al.(2018)Misra, Bhardwaj, and Mishra]{misra2018local}
V.~Misra, A.~Bhardwaj, and A.~Mishra.
\newblock Local onset and demise of the indian summer monsoon.
\newblock \emph{Clim. Dyn.}, 51:\penalty0 1609--1622, 2018.

\bibitem[Moncrieff(2010)]{moncrieff2010multiscale}
M.~W. Moncrieff.
\newblock The multiscale organization of moist convection and the intersection
  of weather and climate.
\newblock \emph{Climate dynamics: Why does climate vary}, 189:\penalty0 3--26,
  2010.

\bibitem[Moncrieff et~al.(2012)Moncrieff, Waliser, Miller, Shapiro, Asrar, and
  Caughey]{moncrieff2012multiscale}
M.~W. Moncrieff, D.~E. Waliser, M.~J. Miller, M.~A. Shapiro, G.~R. Asrar, and
  J.~Caughey.
\newblock Multiscale convective organization and the yotc virtual global field
  campaign.
\newblock \emph{Bull. Am. Meteorol. Soc.}, 93\penalty0 (8):\penalty0
  1171--1187, 2012.

\bibitem[Noska and Misra(2016)]{noska2016characterizing}
R.~Noska and V.~Misra.
\newblock Characterizing the onset and demise of the indian summer monsoon.
\newblock \emph{Geophys. Res. Lett.}, 43\penalty0 (9):\penalty0 4547--4554,
  2016.

\bibitem[Pai and Nair(2009)]{pai2009summer}
D.~Pai and R.~M. Nair.
\newblock Summer monsoon onset over kerala: New definition and prediction.
\newblock \emph{J. Earth Syst. Sci.}, 118\penalty0 (2):\penalty0 123--135,
  2009.

\bibitem[Pai et~al.(2020)Pai, ARTI, SUNITHA, MADHURI, Badwaik, Kundale,
  SULOCHANA, Mohapatra, and Rajeevan]{pai2020normal}
D.~Pai, B.~ARTI, D.~SUNITHA, M.~MADHURI, M.~Badwaik, A.~Kundale, G.~SULOCHANA,
  M.~Mohapatra, and M.~Rajeevan.
\newblock Normal dates of onset/progress and withdrawal of southwest monsoon
  over india.
\newblock \emph{Mausam}, 71\penalty0 (4):\penalty0 553--570, 2020.

\bibitem[Rajesh and Goswami(2020)]{rajesh2020four}
P.~Rajesh and B.~Goswami.
\newblock Four-dimensional structure and sub-seasonal regulation of the indian
  summer monsoon multi-decadal mode.
\newblock \emph{Clim. Dyn.}, 55\penalty0 (9):\penalty0 2645--2666, 2020.

\bibitem[Rani et~al.(2021)Rani, Arulalan, George, Rajagopal, Renshaw, Maycock,
  Barker, and Rajeevan]{rani2021imdaa}
S.~I. Rani, T.~Arulalan, J.~P. George, E.~Rajagopal, R.~Renshaw, A.~Maycock,
  D.~M. Barker, and M.~Rajeevan.
\newblock Imdaa: High-resolution satellite-era reanalysis for the indian
  monsoon region.
\newblock \emph{J. Clim.}, 34\penalty0 (12):\penalty0 5109--5133, 2021.

\bibitem[Saha et~al.(2023)Saha, Mahanta, and Goswami]{saha2023present}
P.~Saha, R.~Mahanta, and B.~Goswami.
\newblock Present and future of the south asian summer monsoon’s rainy season
  over northeast india.
\newblock \emph{npj Clim. Atmos. Sci.}, 6\penalty0 (1):\penalty0 170, 2023.

\bibitem[Sanap et~al.(2019)Sanap, Priya, Sawaisarje, and
  Hosalikar]{sanap2019heavy}
S.~Sanap, P.~Priya, G.~Sawaisarje, and K.~Hosalikar.
\newblock Heavy rainfall events over southeast peninsular india during
  northeast monsoon: Role of el ni{\~n}o and easterly wave activity.
\newblock \emph{Int. J. Climatol.}, 39\penalty0 (4):\penalty0 1954--1968, 2019.

\bibitem[Tomas and Webster(1997)]{tomas1997role}
R.~A. Tomas and P.~J. Webster.
\newblock The role of inertial instability in determining the location and
  strength of near-equatorial convection.
\newblock \emph{Q. J. R. Meteorol. Soc}, 123\penalty0 (542):\penalty0
  1445--1482, 1997.

\bibitem[Walker and Bordoni(2016)]{walker2016onset}
J.~M. Walker and S.~Bordoni.
\newblock Onset and withdrawal of the large-scale south asian monsoon: A
  dynamical definition using change point detection.
\newblock \emph{Geophys. Res. Lett.}, 43\penalty0 (22):\penalty0 11--815, 2016.

\bibitem[Wang(2006)]{wang2006asian}
B.~Wang.
\newblock \emph{The Asian Monsoon}.
\newblock Springer Science \& Business Media, 2006.

\bibitem[Wang et~al.(2009)Wang, Ding, and Joseph]{wang2009objective}
B.~Wang, Q.~Ding, and P.~Joseph.
\newblock Objective definition of the indian summer monsoon onset.
\newblock \emph{J. Clim.}, 22\penalty0 (12):\penalty0 3303--3316, 2009.

\bibitem[Wang et~al.(2002)]{wang2002rainy}
B.~Wang et~al.
\newblock Rainy season of the asian--pacific summer monsoon.
\newblock \emph{J. Clim.}, 15\penalty0 (4):\penalty0 386--398, 2002.

\bibitem[Webster et~al.(1998)Webster, Magana, Palmer, Shukla, Tomas, Yanai, and
  Yasunari]{webster1998monsoons}
P.~J. Webster, V.~O. Magana, T.~Palmer, J.~Shukla, R.~Tomas, M.~Yanai, and
  T.~Yasunari.
\newblock Monsoons: Processes, predictability, and the prospects for
  prediction.
\newblock \emph{J. Geophys. Res. Oceans}, 103\penalty0 (C7):\penalty0
  14451--14510, 1998.

\bibitem[Xavier et~al.(2007)Xavier, Marzin, and Goswami]{xavier2007objective}
P.~K. Xavier, C.~Marzin, and B.~N. Goswami.
\newblock An objective definition of the indian summer monsoon season and a new
  perspective on the enso--monsoon relationship.
\newblock \emph{Q. J. R. Meteorol. Soc}, 133\penalty0 (624):\penalty0 749--764,
  2007.

\end{thebibliography}
\bibliographystyle{abbrvnat}

\subsection*{Acknowledgments}
R.I.S. acknowledges the funding from the Institutions of Eminence (IOE) initiative \\(No. SP22231222CPETWOCTSHOC). B.N.G. thanks Gauhati University for support through the Honorary Professor of Excellence Fellowship. S.T. acknowledges the support from the Prime Minister Research Fellowship, Government of India. The authors are grateful to Professor Rajarshi Roy (University of Maryland, USA) for fruitful suggestions on improving the readability of the manuscript.
\subsection*{Author Contributions}
R.I.S., G.C. and S.T. conceived the study. Y.P. reviewed the literature with inputs from all authors and assembled datasets. G.C. and S.T. designed the network construction methodology, and Y.P. implemented it. B.N.G. conceptualized the large-scale monsoon onset definition. Y.P. visualized the data and produced the figures with inputs from all the authors. All authors contributed to the analysis and interpretation. B.N.G., Y.P. and G.C. wrote the manuscript. 
S.T. and R.I.S. reviewed and edited the manuscript.
\subsection*{Competing Interests}
The authors declare that there are no competing interests.
\subsection*{Additional Information}
\textbf{Supplementary Information} is available for this paper.
\newline
\textbf{Correspondence and requests for materials} should be addressed to R.I.S.

\newpage
\section*{Extended Data figures and tables}

\begin{extdatafigure}[ht!]
    \centering
    \includegraphics[width=\textwidth]{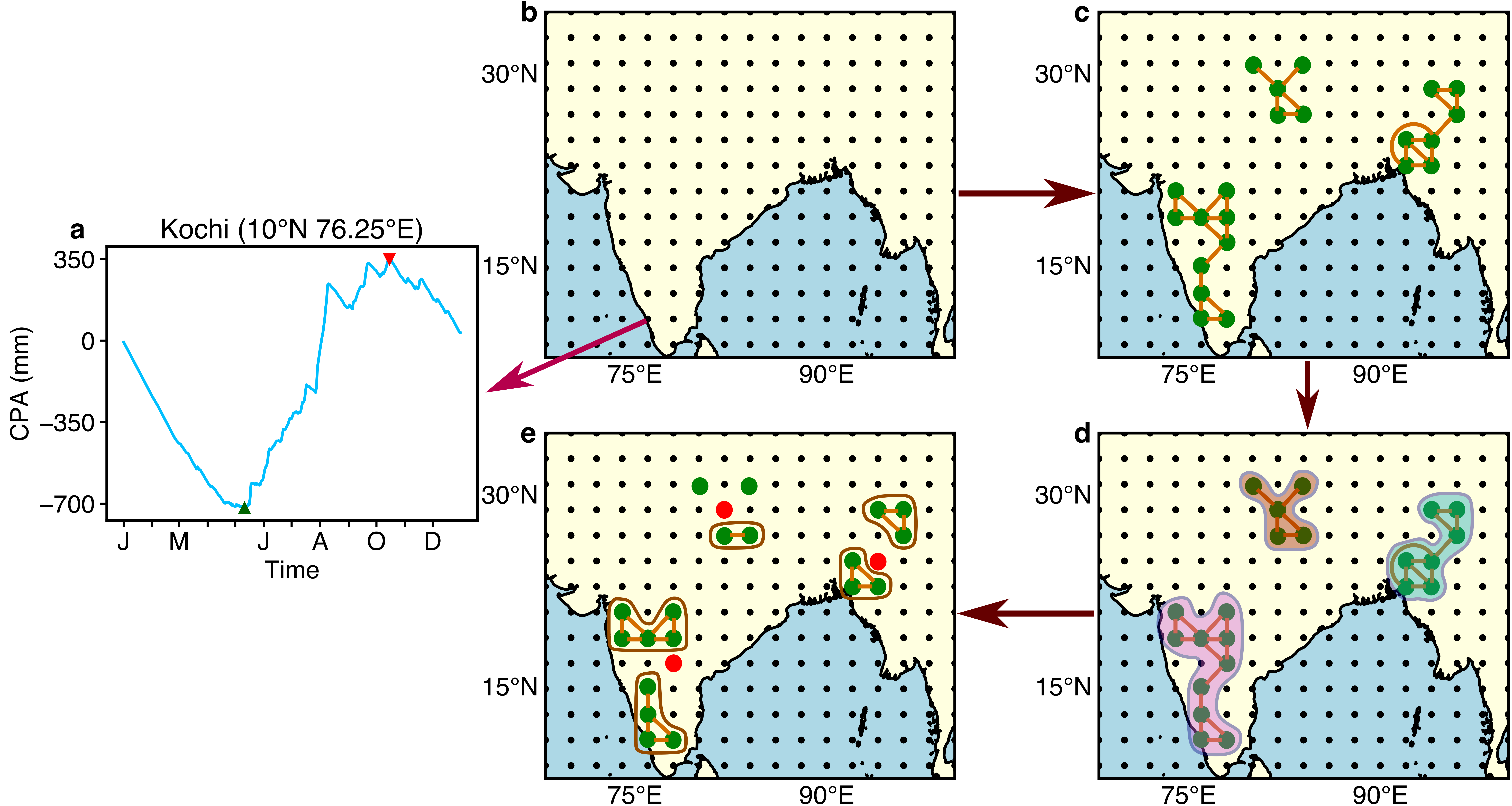}
    \caption{\doublespacing\textbf{Schematic of the procedure for network construction and analysis of the connected components of the network on a specific day.} The blue curve in (a) displays the cumulative precipitation anomaly (CPA) at a specific location, Kochi ($10^\circ$N, 76.25$^\circ$E) for the year 2020. The minima ($\blacktriangle$) and maxima ($\blacktriangledown$) of the CPA curve (blue) indicate the local onset and withdrawal at that location, respectively. We first determine the local onset and withdrawal at each location shown in (b). We establish links between two nodes only if both nodes belong to the Moore neighbourhood and have undergone local onset in (c). In a 2D square lattice, the Moore neighbourhood consists of a central cell along with the eight cells surrounding it. Connecting all the nodes in the Moore neighbourhood except the central node enforces the condition of spatial proximity in the network. During the onset phase in (d), we observe the formation of connected components in the network topology, where the first three largest components are coloured in light shades of magenta, green, and brown, respectively. The largest component, called the giant component (GC), is coloured magenta. Once a node undergoes local withdrawal (marked in red), as shown in (e), the corresponding links are removed from the network, and the connected components disintegrate. The spatial proximity network is updated each day throughout a given year, thus making it a time-varying spatial proximity network.
    } 
    \label{fig: schematic}
\end{extdatafigure}
\newpage
\begin{extdatafigure}[ht!]
    \centering
    \includegraphics[width=1\textwidth]{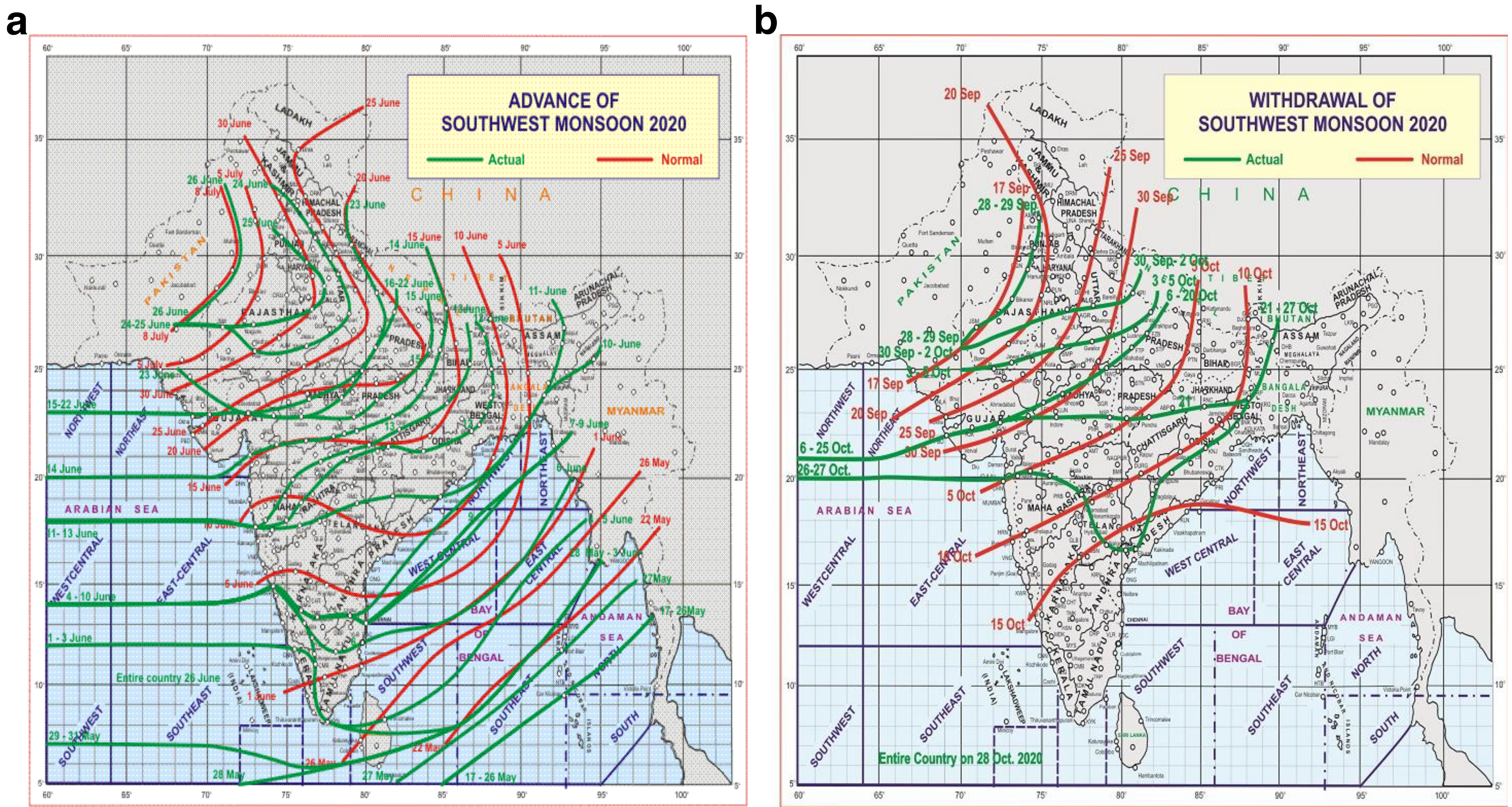}
    \caption{\doublespacing\textbf{Isochrones of (a) onset, and (b) demise of the Indian summer monsoon (ISM) as defined by the India Meteorological Department (IMD) reproduced from IMD Report (2020)}. The IMD compares the actual (green) isochrones of monsoon onset and withdrawal for every year with that of the climatological isochrones (red). However, these climatological isochrones themselves suffer from great uncertainty as they are based on local rainfall\citep{pai2020normal}. Therefore, these isochrones should not be used to determine the interannual variations in the onset and withdrawal of monsoon.}
    \label{fig: imd_isochrones_2020}
\end{extdatafigure}
\newpage
\begin{extdatafigure}[ht!]
    \centering
    \includegraphics[width=1\textwidth]{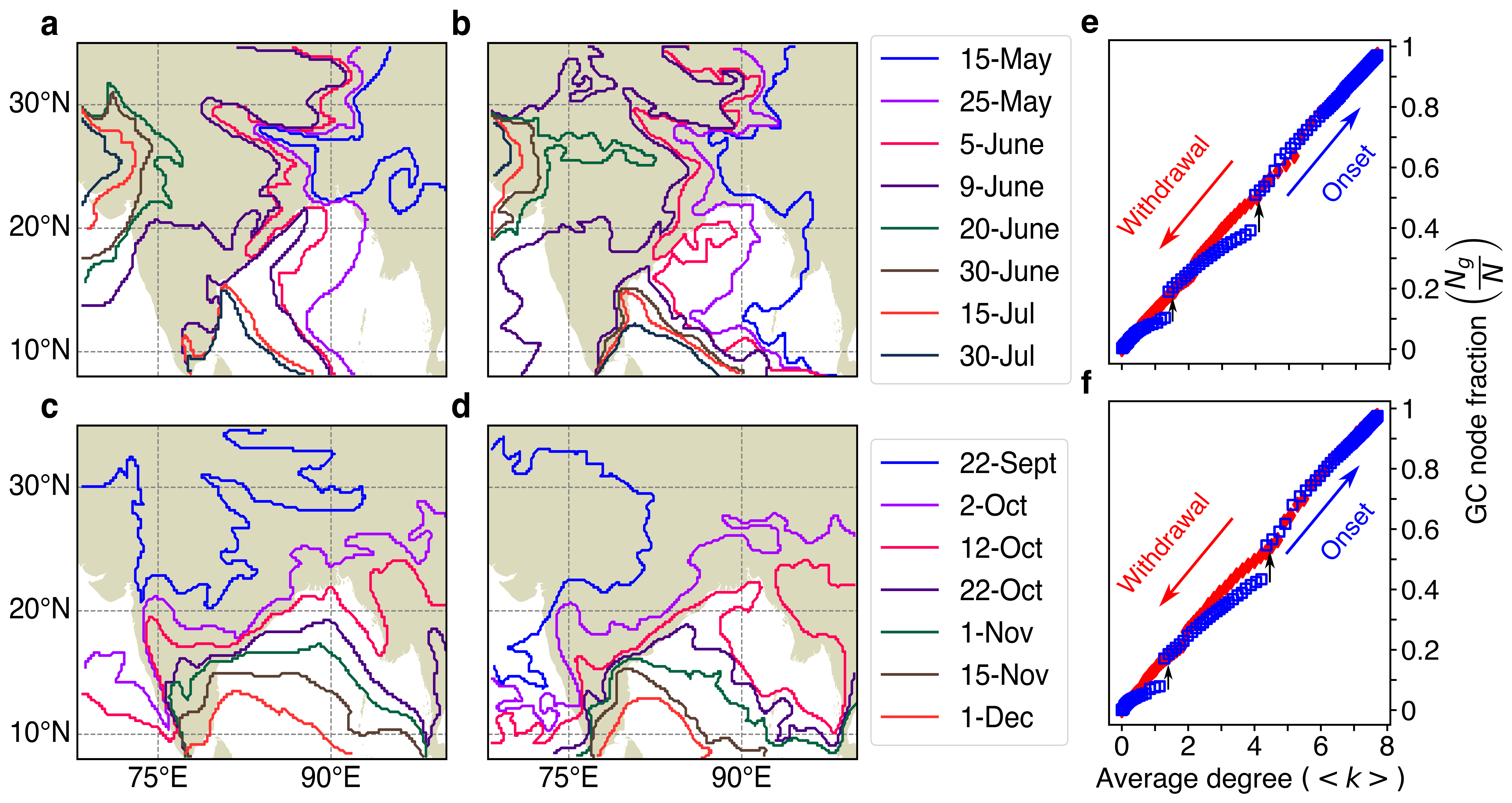}
    \caption{\doublespacing\textbf{Comparison of monsoon isochrones obtained from our network based approach using other precipitation datasets.} Isochrones of large-scale onset (a, b), and large-scale withdrawal (c, d), using the IMDAA precipitation dataset (a, c) and IMERG precipitation dataset (b, d). The network phase diagram obtained using the IMDAA precipitation dataset (e) and IMERG precipitation dataset (f). The two abrupt transitions (black arrows in (e) and (f)) over NEI and MoK are consistent across both precipitation datasets and ERA5 data in the Main text (Fig.\ref{fig: components}m)}
    \label{fig: isochrones_all_datasets}
\end{extdatafigure}
\newpage
\begin{extdatafigure}[ht!]
    \centering
    \includegraphics[width=1\textwidth]{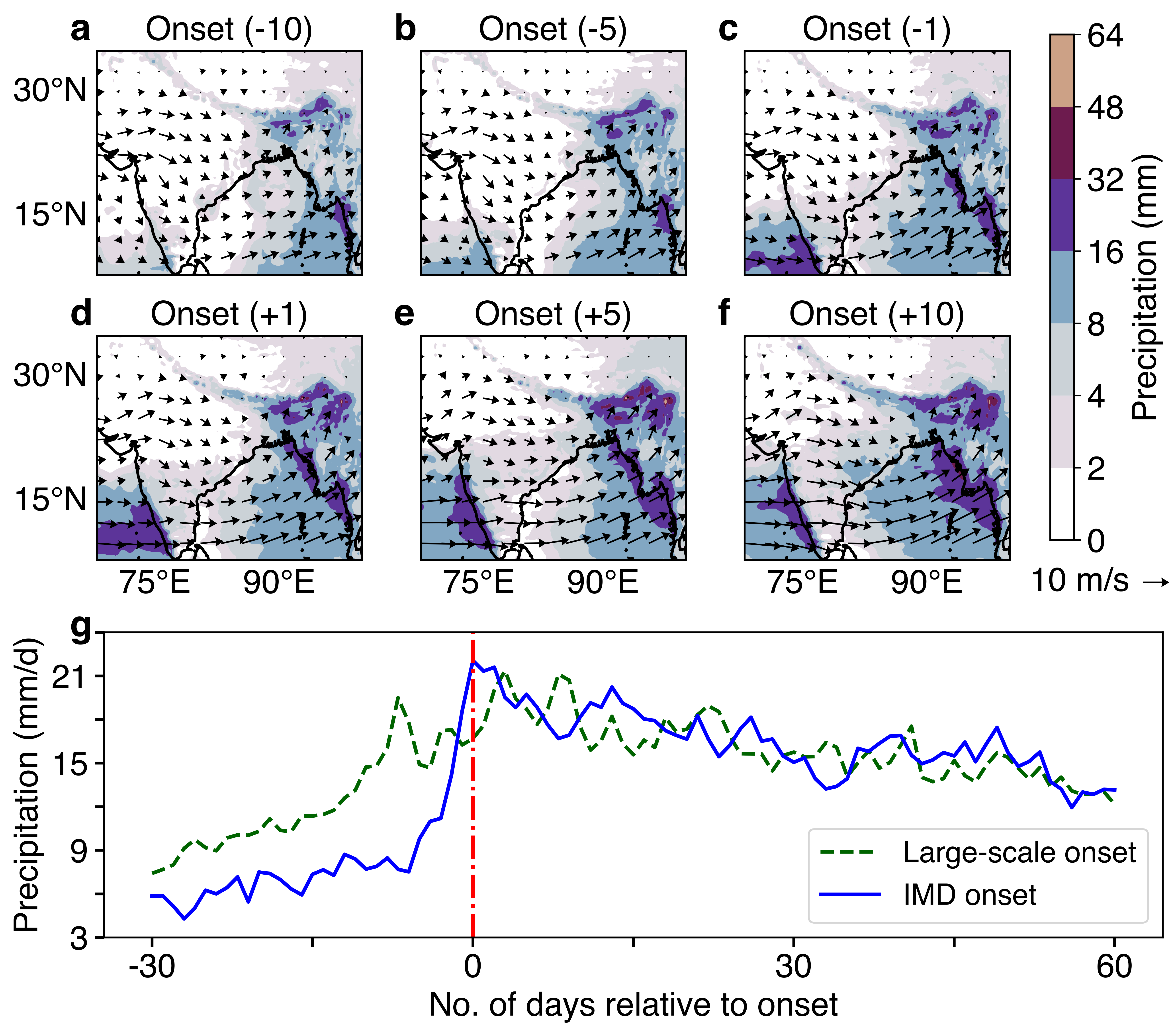}
    \caption{\doublespacing\textbf{Lead-lag composites of wind and precipitation from 1960 - 2023 centred w.r.t the IMD onset dates} over the 14 stations in Kerala used by IMD to declare the MoK. The temporal average of wind and precipitation at a lag of (a) 10 days, (b) 5 days, (c) 1 day, and lead of (d) 1 day, (e) 5 days, (f) 10 days with respect to the MoK defined by IMD. The precipitation over the southwest tip of the peninsula in (d) is substantial but lacks organisation, indicating the dominance of synoptic activity. This approach fails to represent the large-scale monsoon winds, unlike that in the Main text (Fig.3d). A comparison of the spatially averaged precipitation over the 14 stations in Kerala from IMD and large-scale MoK is shown in (g). The local rainfall over Kerala increases slowly from 9 mm/day to 18 mm/day in 30 days using the large-scale onset method, while that by the IMD doubles in 10 days around the onset.} 
    \label{fig: IMD_composites}
\end{extdatafigure}
\newpage
\begin{extdatafigure}[ht!]
    \centering
    \includegraphics[width=1\textwidth]{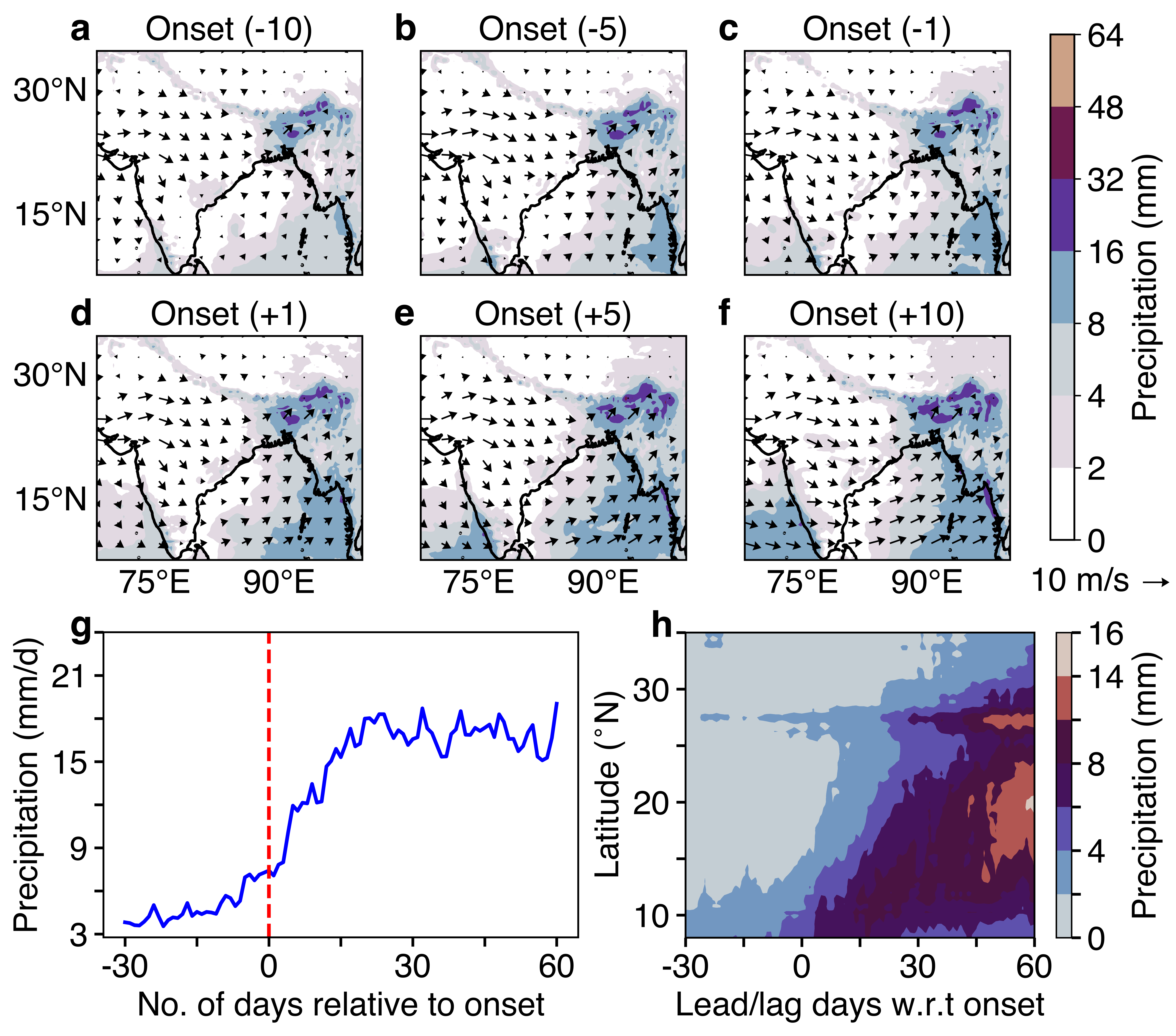}
    \caption{\doublespacing\textbf{Lead-lag composites of wind and precipitation from 1940 - 2023 centred w.r.t the local scale onset date\citep{misra2018local}} over the 14 stations in Kerala. The temporal average of wind and precipitation at a lag of (a) 10 days, (b) 5 days, (c) 1 day, and lead of (d) 1 day, (e) 5 days, (f) 10 days with respect to the MoK. There is no organised precipitation or winds over the peninsula in (a)-(f). The spatially averaged precipitation over the 14 stations in Kerala is shown in (g). The precipitation continues to increase even after 15 days of the onset and then keeps fluctuating. Hovm\"oller diagram for the zonally averaged precipitation between $70^\circ$E - $90^\circ$E is shown in (h). These onsets are triggered mainly by a MISO episode that can produce rainfall over the southern peninsula. However, these MISOs are not ready to propagate northward and are, therefore, prone to the detection of false onsets. The precipitation increases gradually post 15 days of the local onset.} 
    \label{fig: local_composites}
\end{extdatafigure}
\newpage
\begin{extdatafigure}[ht!]
    \centering
    \includegraphics[width=1\textwidth]{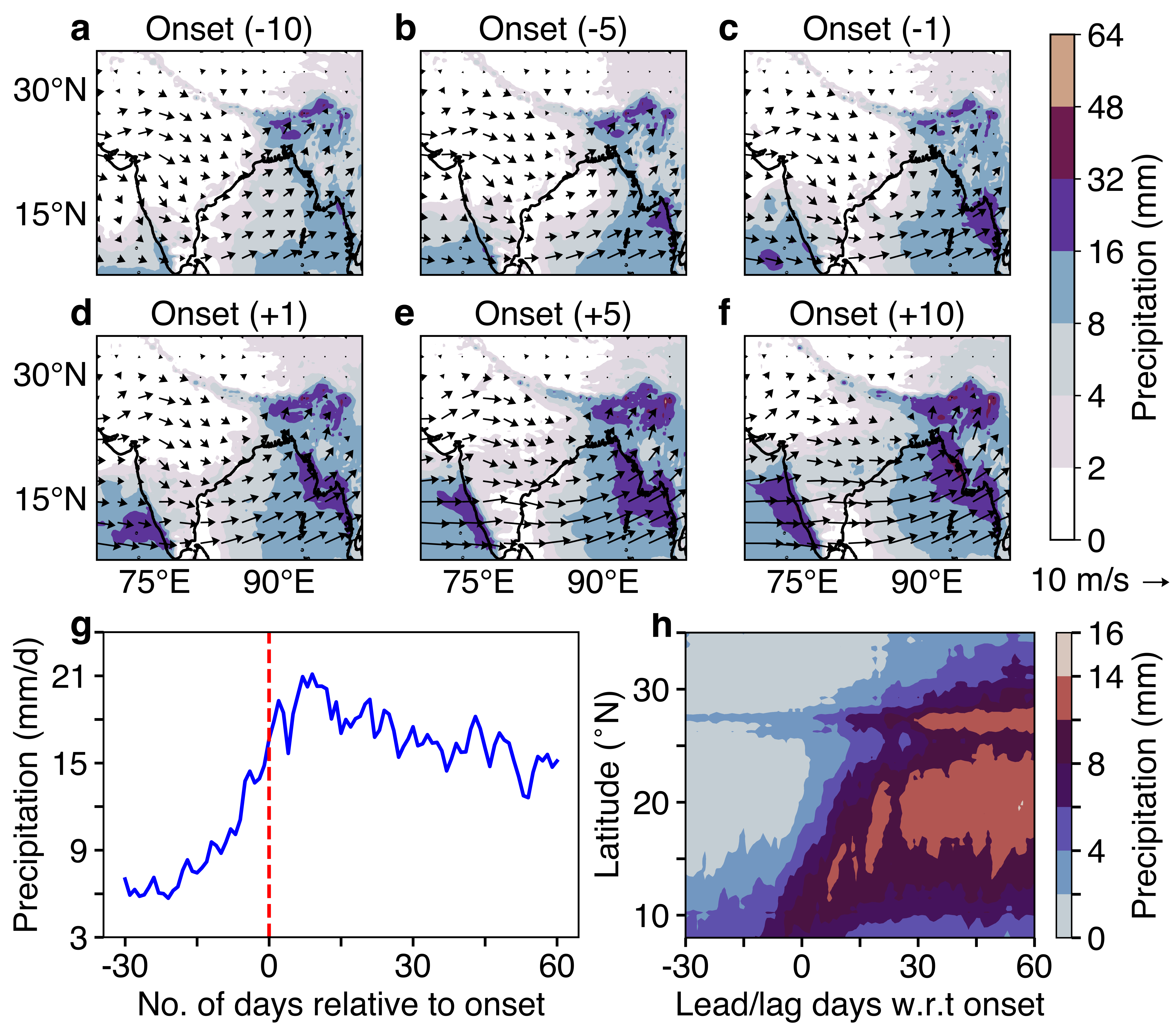}
    \caption{\doublespacing\textbf{Lead-lag composites of wind and precipitation from 1940 - 2022 centred w.r.t regional Monsoon onset over Kerala (MoK) determined from tropospheric temperature (TT) gradient\citep{xavier2007objective}:} The temporal average of wind and precipitation at a lag of (a) 10 days, (b) 5 days, (c) 1 day, and lead of (d) 1 day, (e) 5 days, (f) 10 days with respect to the MoK. A weak organisation of winds and precipitation over the Arabian Sea is seen in (d)-(f) post the onset. The spatially averaged precipitation over the 14 stations in Kerala is shown in (g). Hovm\"oller diagram for the zonally averaged precipitation between $70^\circ$E - $90^\circ$E is shown in (h). The two precipitation bands, after the TT onset, propagate to approximately $16^\circ$N. The northward propagation of MISO is established only after 20 days of TT onset, which makes this method prone to the determination of bogus onsets.}
    \label{fig: TT_MoK_composites}
\end{extdatafigure}
\newpage
\begin{extdatafigure}[ht!]
    \centering
    \includegraphics[width=1\textwidth]{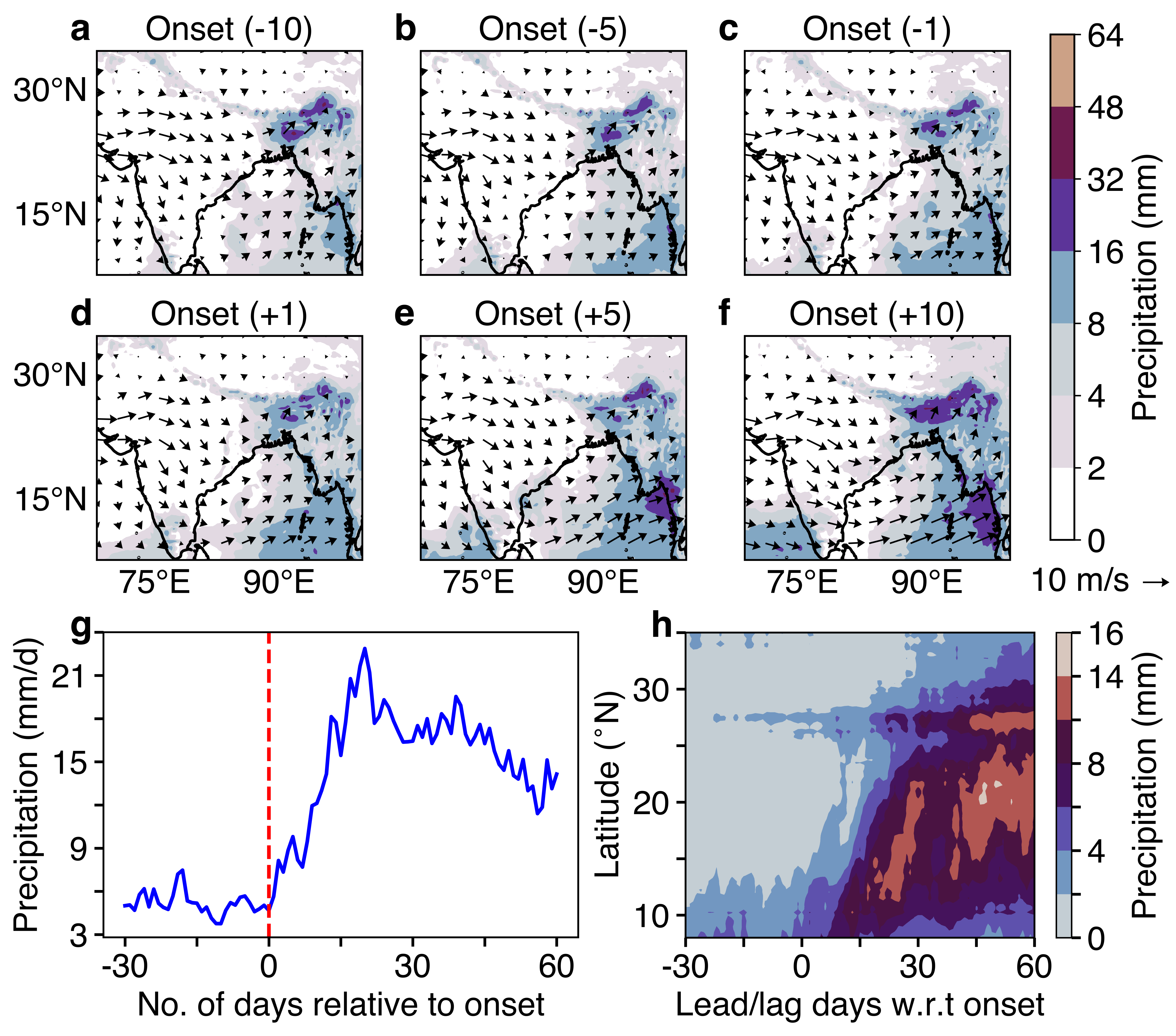}
    \caption{\doublespacing\textbf{Lead-lag composites of wind and precipitation from 1980 - 2015 w.r.t regional Monsoon onset over Kerala (MoK) determined from Change Point (CHP)\citep{walker2016onset} Index:} The temporal average of wind and precipitation at a lag of (a) 10 days, (b) 5 days, (c) 1 day, and lead of (d) 1 day, (e) 5 days, (f) 10 days with respect to the MoK. There is no organisation of winds and precipitation even after 10 days of the onset in (f). The spatially averaged precipitation of the 14 stations in Kerala shown in (g) continues to increase up to 15 days after the onset, after which it starts reducing. Hovm\"oller diagram for the zonally averaged precipitation between $70^\circ$E - $90^\circ$E is shown in (h). The northward propagation of rainfall is seen only after 20 days post the onset. This definition is prone to the identification of false onsets.}
    \label{fig: CHP_MoK_composites}
\end{extdatafigure}
\newpage
\begin{extdatafigure}[ht!]
    \centering
    \includegraphics[width=1\textwidth]{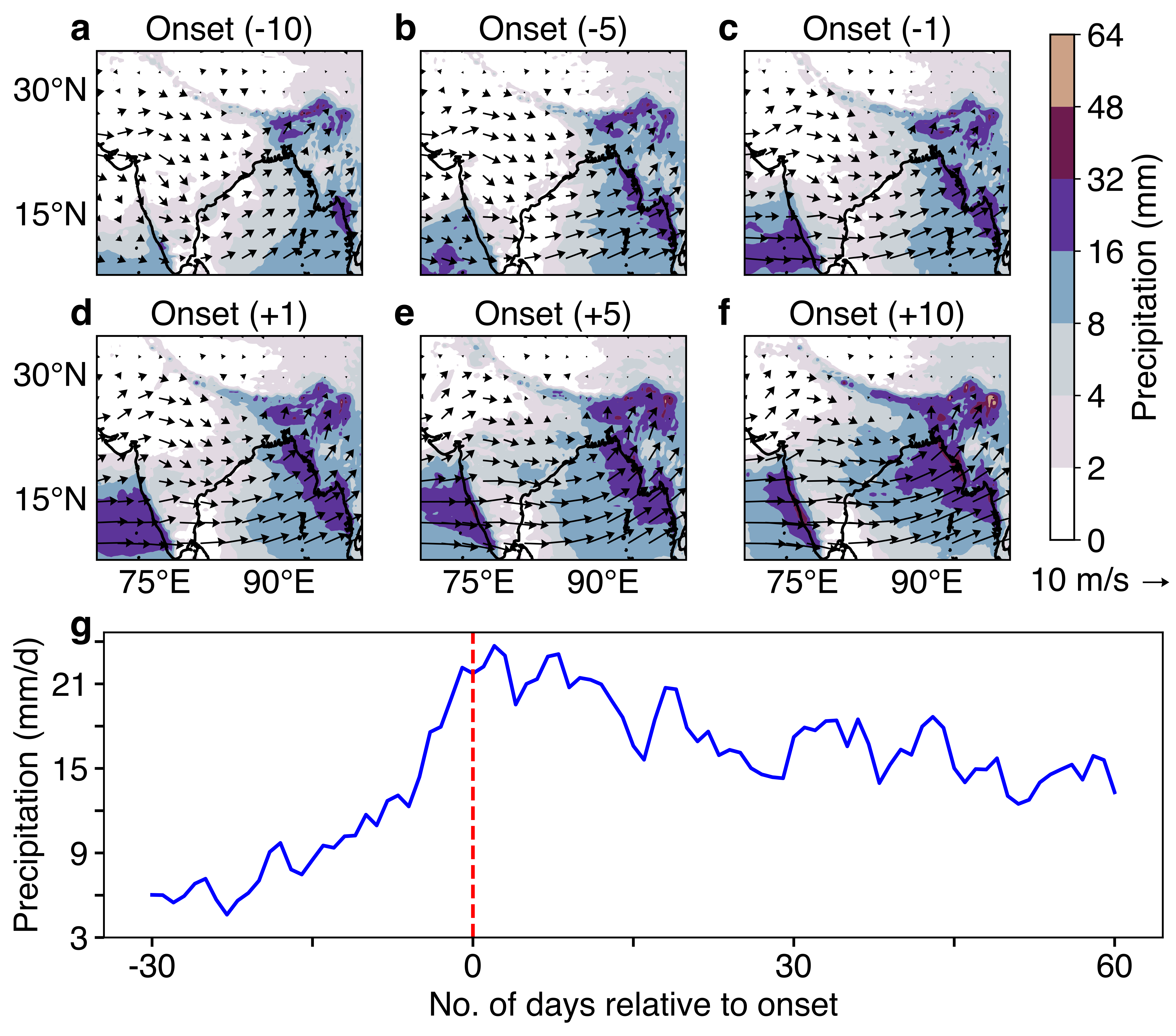}
    \caption{\doublespacing\textbf{Lead-lag composites of wind and precipitation from 1948 - 2000 w.r.t regional Monsoon onset over Kerala (MoK) determined from Hydrological Onset and Withdrawal (HOWI)\citep{fasullo2003hydrological} Index:} The temporal average of wind and precipitation at a lag of (a) 10 days, (b) 5 days, (c) 1 day, and lead of (d) 1 day, (e) 5 days, (f) 10 days with respect to the MoK. The precipitation and winds are well-organised over the Arabian Sea. The spatially averaged precipitation over the 14 stations in Kerala is shown in (g). The precipitation reaches a maximum on the onset day, after which it starts fluctuating and is then sustained after 30 days of the onset.}
    \label{fig: HOWI_MoK_composites}
\end{extdatafigure}
\newpage
\begin{extdatafigure}[ht!]
    \centering
    \includegraphics[width=1\textwidth]{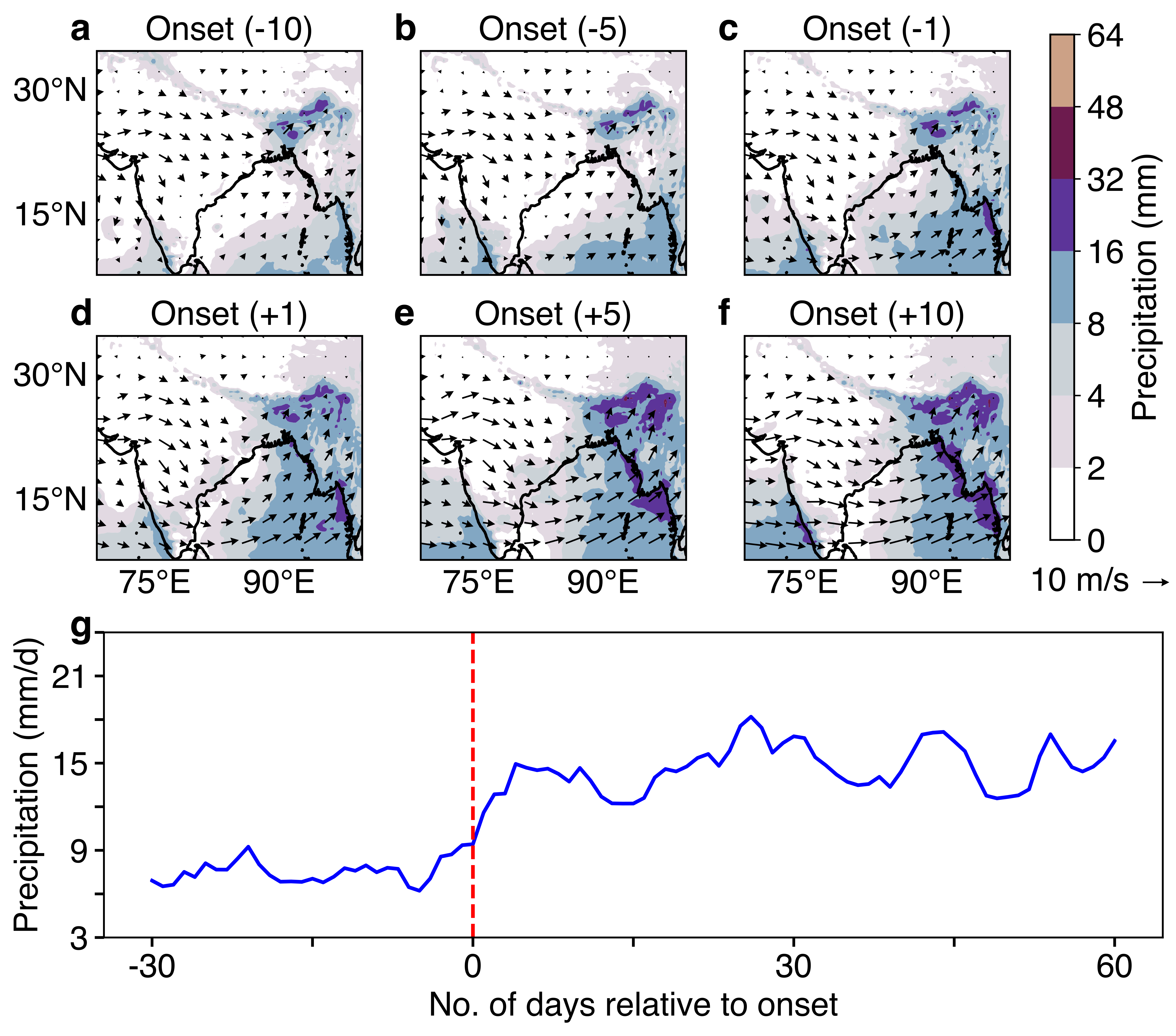}
    \caption{\doublespacing\textbf{Lead-lag composites of 850 hPa winds and precipitation from 1940 - 2023 w.r.t network-based large-scale onset dates centred around Northeast India:} ($22^\circ$N - $26^\circ$N, $90^\circ$E - $94^\circ$E) The temporal average of wind and precipitation at a lag of (a) 10 days, (b) 5 days, (c) 1 day, and lead of (d) 1 day, (e) 5 days, (f) 10 days with respect to the NEI onset. The organised wind and precipitation over the NEI are consistent with the low-level cyclonic vortex as proposed by \citet{das2024dynamics} The spatially averaged precipitation over Northeast India is shown in (g). The large-scale onset over NEI occurs abruptly as precipitation increases threefold times in approximately two weeks.}
    \label{fig: NEI_composites}
\end{extdatafigure}
\newpage
\begin{extdatafigure}[ht!]
    \centering
    \includegraphics[width=1\textwidth]{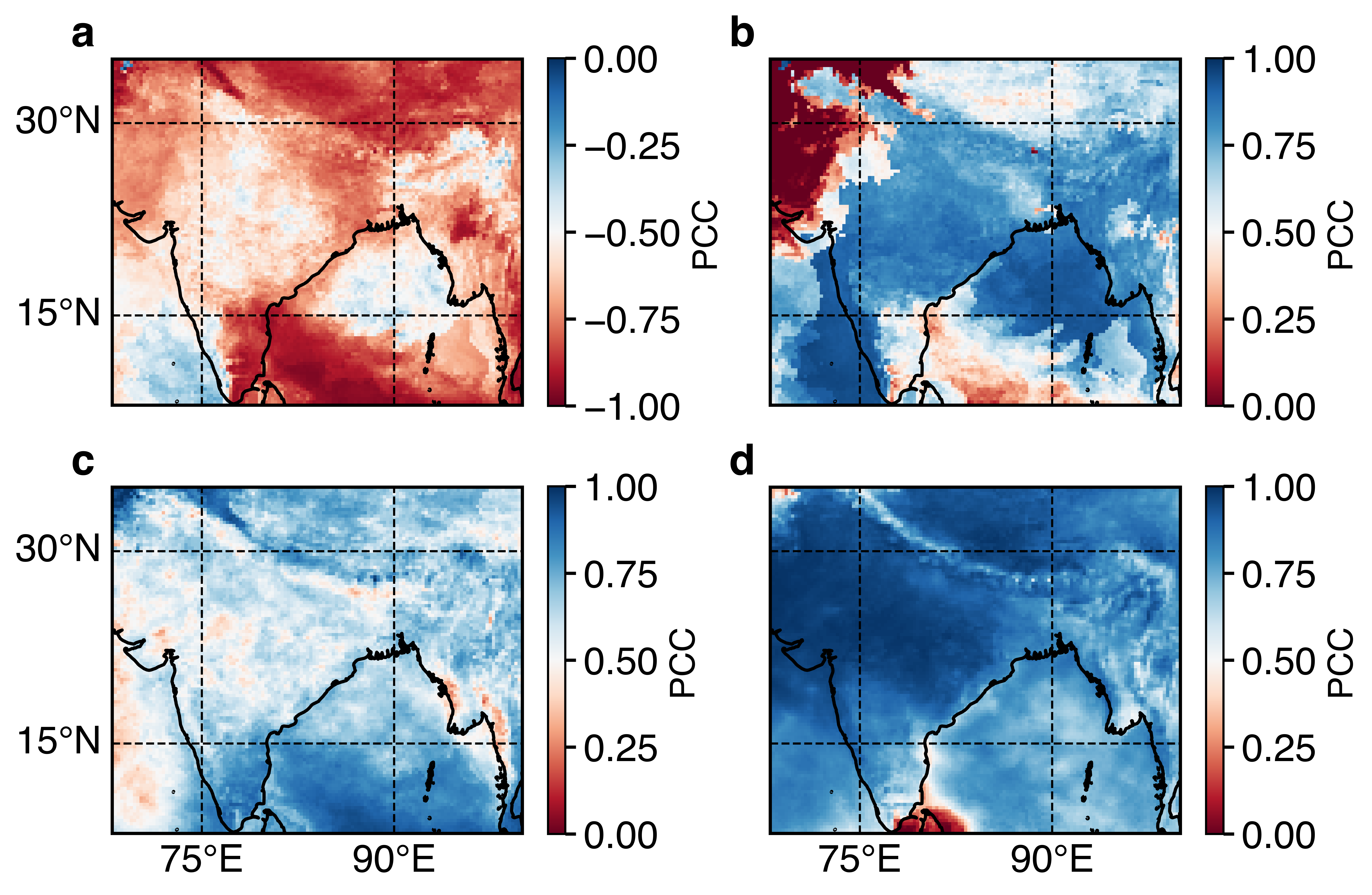}
    \caption{\doublespacing\textbf{Pearson correlation coefficient (PCC) of LRS accumulated precipitation with:} the large-scale (a) onset and (b) withdrawal dates, (c) large-scale LRS, and (d) June-September (JJAS) precipitation. The negative correlation in (a) signifies that the accumulated LRS precipitation is less if the onset is delayed and vice-versa. The positive correlation in (b) and (c) indicates an increase in accumulated LRS precipitation with the delay in withdrawal and increase in LRS, respectively. The LRS accumulated precipitation has a significant correlation with the JJAS precipitation in (d).}
    \label{fig: LRS_prep_correlation}
\end{extdatafigure}

\end{document}